\documentclass[prl,twocolumn,superscriptaddress]{revtex4}
\usepackage{amsmath,amssymb,graphicx,color,bm,epstopdf}

\begin{document}

\title{Superradiant Terahertz Emission by Dipolaritons}
\author{O. Kyriienko}
\affiliation{Science Institute, University of Iceland, Dunhagi-3, IS-107, Reykjavik,
Iceland}
\affiliation{Division of Physics and Applied Physics, Nanyang Technological University
637371, Singapore}
\author{A. V. Kavokin}
\affiliation{School of Physics and Astronomy, University of Southampton, Highfield,
Southampton SO17 1BJ, United Kingdom}
\affiliation{Spin Optics Laboratory, St. Petersburg State University, 1, Ulianovskaya,
198504, Russia}
\author{I. A. Shelykh}
\affiliation{Science Institute, University of Iceland, Dunhagi-3, IS-107, Reykjavik,
Iceland}
\affiliation{Division of Physics and Applied Physics, Nanyang Technological University
637371, Singapore}
\date{\today}

\begin{abstract}
Dipolaritons are mixed light-matter quasiparticles formed in double quantum wells embedded in microcavities. Due to resonant coupling between direct and indirect excitons via electronic tunnelling, dipolaritons possess large dipole moments. Resonant excitation of the cavity mode by a short pulse of light induces oscillations of the indirect exciton density with a characteristic frequency of Rabi flopping. This results in oscillations of classical Hertz dipoles array which generate supperradiant emission on a terahertz (THz) frequency. Resulting THz signal may be enhanced using the supplementary THz cavity in the weak coupling regime.
\end{abstract}

\maketitle


\emph{Introduction.---}The generation and frequency modulation of terahertz radiation are among the major technological challenges nowadays \cite{Siegel}. Applications of terahertz sources span from communication technologies to medicine and security. The existing THz emitters are based on a large variety of physical principles \cite{Gallerano}. One possible solution is a
conventional solid state oscillator based on a high frequency Gunn or tunnelling diode \cite{Eisele}. The operating frequency of such oscillator is limited to the microwave region and the lower boundary of terahertz range. Second, quantum cascade lasers (QCL) allow for coherent emission of radiation in a terahertz range \cite{Faist,Kohler}. Exploiting the multiple photon emission from intersubband transition of wide quantum wells \cite{Geiser}, QCL cover the upper boundary of THz range with relatively high power of emission (up to 50 mW). QCL operate at cryogenic temperatures which strongly limits their application area. The laser driven terahertz emitters form a third group of THz sources, where femtosecond optical pulse illuminating the semiconductor structure leads to oscillation of carrier density which generates terahertz radiation \cite{Auston,Fattinger,Heinz,Johnston}. Finally, the wide group of THz emitters are free electron based sources like klystron or free electron laser. These are powerful but quite bulky sources. Optimization of spectral characteristics, size, efficiency, cost and operation temperature of terahertz devices is one of the priority objectives of modern optoelectronics. Here we propose a compact optical to terahertz radiation converter allowing for efficient frequency modulation and operating at high temperatures.

Theoretical proposals for THz sources in solid state and semiconductor physics present a rich diversity. One example is the use of carbon nanostructures, in particular, carbon nanotubes (CNT's) \cite{Kibis,Portnoi}, or Aharonov-Bohm quantum rings \cite{Alexeev}. Another area where a possibility of THz generation has been widely studied theoretically is \emph{polaritonics} --- the interdisciplinary research area at the boundary of solid state physics and quantum optics \cite{KavokinBook,PolaritonDevices}. Polaritonic devices are based on the strong coupling between excitons in semiconductor quantum wells and confined photons. Proposals include the polariton based THz emitters where signal is generated from transitions between upper and lower polariton branches \cite{KVKavokin,Savenko,delValle} and transition between $2p$ and $1s$ exciton states, the latter one strongly coupled to the cavity mode \cite{Kavokin}. Recent theoretical studies also
include the proposal for bosonic cascade lasers realized due to the multiple THz photon emission by excitons (exciton-polaritons) confined in a parabolic potential trap \cite{Liew}. Our present proposal is based on the recent experimental realization of dipolaritons --- cavity exciton-polaritons characterized by large dipole moments \cite{Cristofolini}.

The mechanism for terahertz signal generation described in this Letter relies on the beats between spatially direct and indirect excitons. Spatially indirect excitons are composed of electrons and holes situated in separate quantum wells (QWs) \cite{Lozovik}. Coupled QW systems where separation of electrons and holes and energy splitting between direct and indirect excitons can be controlled by applied electric fields have been studied in the recent decades \cite{ButovPRL,SnokeScience,ButovNature}. The radiative lifetimes of indirect excitons are usually much longer than those
of direct excitons due to the lower electron--hole overlap. Another important feature of indirect excitons is their large dipole moments in the normal to QW plane direction resulting in strong exciton-exciton interactions \cite{KyriienkoIndirect}.

Recently, it has been shown that exciton polaritons and spatially indirect excitons can be intermixed in the biased semiconductor microcavities with embedded coupled QWs \cite{ChristmannPRB,ChristmannAPL,Cristofolini}. In these structures new quasiparticles being linear superposition of cavity photon (C), direct exciton (DX) and indirect exciton (IX) appear. They form three exciton-polariton modes, namely, the upper polariton (UP), middle polariton (MP) and lower polariton (LP) modes. These modes may be characterized by large dipole moments which is why they are referred to as \emph{dipolaritons} \cite{Cristofolini}. Here we study theoretically the effect of beats between dipolariton modes due to the tunnel coupling between direct and indirect excitons, which result into superradiant THz emission by array of Hertz dipoles. Its characteristics can be tuned by the applied bias and the calculated efficiency of proposed emitter is comparable with state-of-art generators.

\begin{figure}[t]
\includegraphics[width=0.9\linewidth]{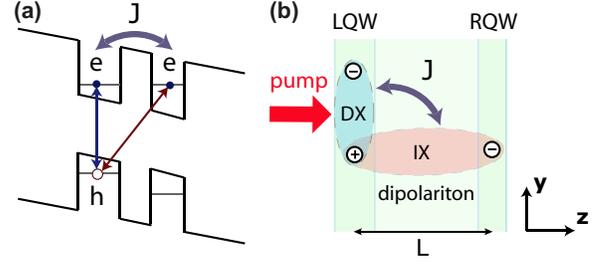}
\caption{(Color online) Sketch of the system. (a) Double quantum well (DQW) heterostructure with resonantly coupled electron levels and a hole in the left quantum well (QW). The right QW has larger bandgap than left QW tuned by the presence of indium alloy or width of the well. (b) Electron-hole bilayer with schematic picture of coupled spatially indirect and direct exciton which form dipolariton.}
\label{Fig1}
\end{figure}

\emph{The model.---}The structure we consider represents two QWs separated by a barrier which is sufficiently thin to allow for resonant electron tunnelling (Fig. \ref{Fig1}). The electron wave function is shared between two QWs in this case. The optical microcavity is tuned to the wider QW exciton resonance (LQW in Fig. \ref{Fig1}), while the right QW remains decoupled from the optical pump. Pulsed pumping creates electron-hole pairs which form direct excitons. The bias applied in growth direction induces mixing of direct and indirect exciton states. Coupling of the cavity photons to direct excitons resulting in appearance of two polariton modes has been extensively studied \cite{KavokinBook,PolaritonDevices}. The coupling and anticrossing of DX-IX resonances tuned by electric field is also a well-known effect in GaAs/AlGaAs QW structures \cite{Kardiff,Bayer,Kyriienko2011}. Both effects combined result in the appearance of dipolaritons studied recently by Cristofolini \textit{et al.} \cite{Cristofolini}.

The correct treatment of a real dipolariton system involves both coherent and decoherent parts for the Hamiltonian, $\hat{\cal{H}}=\hat{\cal{H}}_{coh}+\hat{\cal{H}}_{dec}$. The Hamiltonian corresponding to the coherent part reads
\begin{align}
\label{Hamiltonian}
\hat{\cal{H}}_{coh}=&\hbar \omega_{C}\hat{a}^{\dagger}\hat{a}+\hbar \omega_{DX}\hat{b}^{\dagger}\hat{b}+\hbar \omega_{IX}\hat{c}^{\dagger}\hat{c}+\frac{\hbar \Omega}{2}(\hat{a}^{\dagger}\hat{b}+\hat{b}^{\dagger}\hat{a})\\ \notag
&-\frac{\hbar J}{2}(\hat{b}^{\dagger}\hat{c}+\hat{c}^{\dagger}\hat{b})+P\hat{a}^{\dagger}+P^{*}\hat{a},
\end{align}
where $\hat{a}^{\dagger},~\hat{a}$, $\hat{b}^{\dagger},~\hat{b}$ and $\hat{c}^{\dagger},~\hat{c}$ are creation and annihilation operators for cavity photons, direct excitons and indirect excitons, respectively. Here $\hbar \omega_{C}$, $\hbar \omega_{DX}$ and $\hbar \omega_{IX}$ denote cavity mode, direct exciton and indirect exciton energies, and first three terms of Hamiltonian describe energy of the bare modes. The fourth and fifth terms describe coupling between the modes, where $\Omega$ denotes the coupling constant between photon and direct exciton, and the tunnelling rate corresponding to DX--IX coupling is $J$. The last two terms in Hamiltonian describe the optical pumping of cavity mode with amplitude $P(t)=P_{0}(t)e^{-i\omega_{P}t}$, where $\omega_{P}$ is a pumping frequency. 

The decoherent part of dynamics of the system is mainly governed by the radiative lifetimes of the modes and phonon-scattering processes, which lead to population of thermalized exciton reservoirs \cite{SM}. They can be described by the exciton-phonon interaction Hamiltonian $\hat{\cal{H}}_{int}=\hat{\cal{H}}^{+} + \hat{\cal{H}}^{-}$, where
\begin{align}
\label{H+}
&\hat{\cal{H}}^{+} = D_{DX} \sum_{k} \hat{b}^{\dagger}\hat{b}_{R,k} \hat{d}^{\dagger}_{k} + D_{IX} \sum_{k} \hat{c}^{\dagger}\hat{c}_{R,k} \hat{d}^{\dagger}_{k},\\
&\hat{\cal{H}}^{-} =  D_{DX} \sum_{k} \hat{b}_{R,k}^{\dagger}\hat{b} \hat{d}_{k} + D_{IX} \sum_{k} \hat{c}_{R,k}^{\dagger}\hat{c} \hat{d}_{k},
\label{H-}
\end{align}
correspond to processes with emission ($\hat{d}^{\dagger}_{k}$) and absorption ($\hat{d}_{k}$) of phonon with in-plane momentum $k$, respectively. Here $\hat{b}_{R,k}^{\dagger},~\hat{b}_{R,k},\hat{c}_{R,k}^{\dagger}$ and $\hat{c}_{R,k}$ are creation and annihilation operators for direct and indirect excitons in reservoirs. $D_{DX,IX}$ denote exciton-photon interaction constants. These incoherent processes can be conveniently treated using master equation for the density matrix, $i\hbar \partial \rho/\partial t =[\hat{\cal{H}}_{coh},\rho] +\hat{\cal{L}}\rho^{(dis)} +\hat{\cal{L}}\rho^{(th)}$, where $\hat{\cal{L}}\rho^{(dis)}$ corresponds to standard Lindblad superoperator describing the finite lifetime of the modes \cite{Savenko,Kavokin}, and $\hat{\cal{L}}\rho^{(th)}$ is responsible for phonon-assisted processes (see details of derivation in \cite{SM}).

In the following we shall be interested in the large occupation numbers of the modes, and mean-field approximation can be applied. The dynamics of occupation numbers of the modes can be found from the master equation with operators treated as classical fields \cite{SM}
\begin{align}
\label{aC}
&\partial \langle \hat{a} \rangle/\partial t = -i \Omega \langle \hat{b} \rangle /2 - \gamma_{C} \langle \hat{a} \rangle /2 -i\widetilde{P},\\
\notag
&\partial \langle \hat{b} \rangle/\partial t = i\delta_{\Omega}\langle \hat{b} \rangle -i J \langle \hat{c} \rangle /2 -i \Omega \langle \hat{a} \rangle /2 - \gamma_{DX}\langle \hat{b} \rangle /2 \\
\label{bDX} &+ W \sum_{k} (\langle \hat{b}_{R,k}^{\dagger} \hat{b}_{R,k} \rangle - n_{\text{ph},k})\langle \hat{b} \rangle,\\
\notag
&\partial \langle \hat{c} \rangle/ \partial t = i(\delta_{\Omega}-\delta_{J})\langle \hat{c} \rangle -i J \langle \hat{b} \rangle /2 - \gamma_{IX} \langle \hat{c} \rangle /2 \\
\label{cIX} &+ W \sum_{k} (\langle \hat{c}_{R,k}^{\dagger} \hat{c}_{R,k} \rangle - n_{\text{ph},k})\langle \hat{c} \rangle,\\
\label{bRDX}
&\partial \langle \hat{b}_{R,q}^{\dagger} \hat{b}_{R,q} \rangle/\partial t = -\gamma_{DX} \langle \hat{b}_{R,q}^{\dagger} \hat{b}_{R,q} \rangle + W \sum_{k} \lbrace |\langle \hat{b} \rangle|^2 \times \\ \notag &(\langle \hat{b}_{R,k}^{\dagger} \hat{b}_{R,k} \rangle +1) n_{\text{ph},k} - (|\langle \hat{b} \rangle|^2 +1) \langle \hat{b}_{R,k}^{\dagger} \hat{b}_{R,k} \rangle (n_{\text{ph},k}+1) \rbrace,\\
\label{cRDX}
&\partial \langle \hat{c}_{R,q}^{\dagger} \hat{c}_{R,q} \rangle/\partial t = -\gamma_{IX} \langle \hat{c}_{R,q}^{\dagger} \hat{c}_{R,q} \rangle + W \sum_{k} \lbrace |\langle \hat{c} \rangle|^2 \times \\ \notag &(\langle \hat{c}_{R,k}^{\dagger} \hat{c}_{R,k} \rangle +1) n_{\text{ph},k} - (|\langle \hat{c} \rangle|^2 +1) \langle \hat{c}_{R,k}^{\dagger} \hat{c}_{R,k} \rangle (n_{\text{ph},k}+1) \rbrace,
\end{align}
where $\langle ...\rangle={\rm Tr}\lbrace ...\rho \rbrace$ denotes averaging using the density matrix $\rho$. Here we applied the rotating wave approximation $\langle\hat{a}_{i}\rangle \rightarrow \langle \hat{a}_{i} \rangle e^{-i\omega_{C}t}$, and introduced the photonic $\delta_{\Omega}=\omega_{C}-\omega_{DX}$ and excitonic $\delta_{J}=\omega_{IX}-\omega_{DX}$ detunings. The pumping amplitude reads as $\widetilde{P}(t)=P_{0}(t) e^{-i\Delta t} /\hbar$, where $\Delta =\omega_{P}-\omega_{C}$ denotes detuning of the coherent optical pumping. The damping rates of the modes are defined by parameters $\gamma_{i}=2\pi /\tau_{i}$, $i=C,~DX,~IX$. Typical lifetimes of the modes are $\tau_{C}\approx 3$ ps, $\tau _{DX}\approx 1$ ns and $\tau_{IX}\approx 100$ ns. The reservoir dynamics is governed by scattering rate $W=\delta_{\Delta E} D_{DX}^2$, being proportional to a square of exciton-phonon interaction constant, and $\delta_{\Delta E}$ is a constant accounting for the energy conservation, which is taken to be inverse broadening of exciton states divided by square of Plank constant. For simplicity we consider $D_{DX}=D_{IX}$, assume $N= 4\times 10^5$ reservoir states, and use total scattering rate $W=2$ ps$^{-1}$ \cite{Savenko}. $n_{ph,k}$ is Bose distribution of phonons. 
\begin{figure}[tbp]
\includegraphics[width=1.0\linewidth]{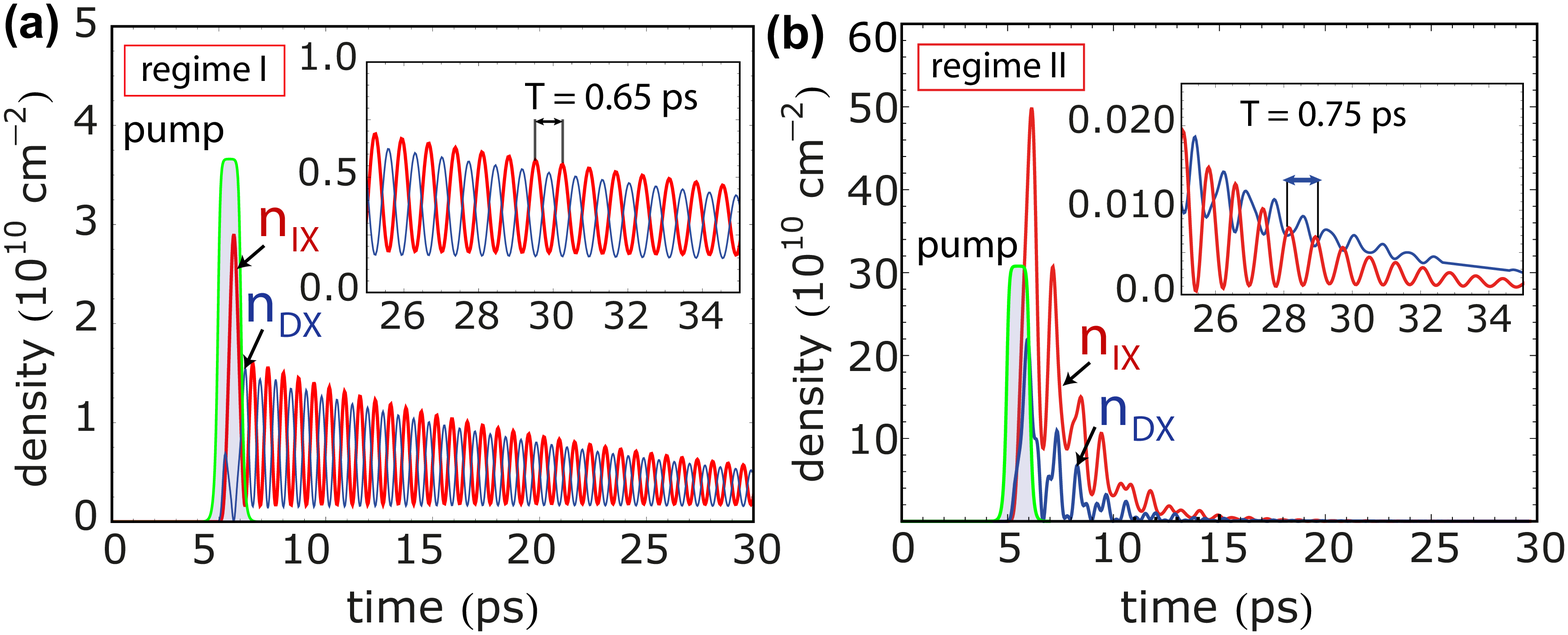}
\caption{(Color online) Dynamics of the dipolariton system subjected to pulsed optical pumping which shows oscillations of indirect exciton ($n_{IX}=|\langle \hat{c} \rangle |^2$) and direct exciton ($n_{DX}=|\langle \hat{b} \rangle |^2$) density. The coupling constants are equal to $\hbar J=\hbar \Omega =6$ meV. (a) Oscillations of density in regime I, where cavity mode is detuned from direct exciton resonance by $\delta_{\Omega}=-10$ meV. Pumping frequency is chosen as $\hbar \Delta =11.5$ meV, and electric field $F=0.95F_{0}$, where $F_{0}$ is field corresponding to IX-DX resonance. The green area schematically represents optical pulse with duration $\Delta \tau =1$ ps (pump, scaled intensity). The inset shows the long-term term oscillations of $n_{IX}(t)$ and $n_{DX}(t)$ density which are in antiphase. The period of oscillations is equal to $T=0.65$ ps. (b) Oscillations corresponding to regime II, where detuning is equal to $\delta_{\Omega}=3$ meV. For the same electrical field and pump intensity ($\hbar \Delta =0$) the magnitude of signal is higher comparing to regime I. Inset shows highly damped long-term oscillations in the second regime.}
\label{Fig2}
\end{figure}

Additionally, we derived dynamic equations for occupation numbers for the modes going to higher order of mean-field theory and verified validity of the system (\ref{aC})--(\ref{bRDX}) for chosen pumping conditions. The former model represents the general description of dipolariton system and allows for accounting of various decoherent processes including pure and phonon-assisted dephasing, which lead to faster decay of oscillations between modes (see \cite{SM}, section B). The effects of non-linearities introduced by exciton-exciton interactions are discussed in the section C of supplemental material.

\textit{Results and discussion.---}We calculate numerically the dynamics of the system with coupled quasiparticles subjected to a picosecond pulsed optical pumping. The presence of mixing terms between different modes implies the oscillating behavior similar to Rabi flopping in the classical model of a two-level system subjected to a time-varying field. It is important to note that our system has several characteristic frequencies. They are governed by the exciton-photon coupling strength $\Omega$ and detuning $\delta_{\Omega}$, the IX-DX coupling strengths $J$ and detuning $\delta_{J}$. Additionally, the pumping frequency $\omega_{P}$ governs the efficiency of the pump. Varying these characteristic frequencies one can control the frequency, magnitude and damping rate of indirect exciton density oscillations.

While the coupling constants $\Omega$ and $J$ are dependent on the geometry of the structure and can hardly be tuned for a given sample, the detunings between modes $\delta_{J}$ and $\delta_{\Omega }$ are strongly sensitive to the applied electric field $F$ and the incidence angle of the cavity pump, respectively \cite{Cristofolini}. Tuning these parameters one can bring the oscillating dipole system to different regimes. If the cavity mode is far-detuned from IX-DX anti-crossing the light-exciton coupling is weak, which will be referred to as the regime I. If the detuning $\delta_{\Omega}$ is small the strong intermixing of IX, DX, and C modes takes place, which corresponds to the regime II.

First, we assume low temperature of the system, when phonon processes are suppressed. The behavior of the system in the regime I is shown in Fig. \ref{Fig2}(a) for the detuning $\delta_{\Omega}=-10$ meV. We observe antiphase oscillations of IX and DX densities with decaying amplitudes. The inset in Fig. \ref{Fig2}(a) shows a zoom of long-standing phase-locked oscillations which last for several tens of picoseconds. For the electric field corresponding to the resonance between IX and DX modes, $F_{0}$, the frequency of oscillations is given by $\nu \approx J/2\pi =1.45$ THz. The electric field at resonance is equal to $F_{0}=\delta_{I-D}/eL$, where $\delta_{I-D}=E_{IX}^{(0)}-E_{DX}^{(0)}$ is exciton detuning at zero applied field, and $L$ is a separation between QWs centers. Considering In$_{0.1}$Ga$_{0.9}$As$/$GaAs$/$In$_{0.08}$Ga$_{0.92}$As (10 nm/4 nm/10 nm) structure studied in Ref. \cite{ChristmannAPL}, where $\delta_{I-D}=18$ meV and $L=14$ nm, it can be calculated as $F_{0}\approx 12.8$ kV/cm.

\begin{figure}[tbp]
\includegraphics[width=1.0\linewidth]{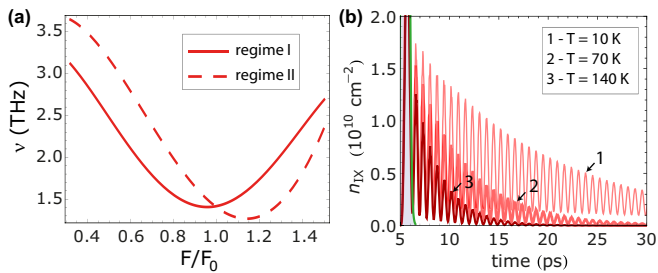}
\caption{(Color online) (a) Frequency of indirect exciton density oscillations as a function of dimensionless electric field calculated for oscillations in regime I (solid line) and regime II (dashed line). (b) Indirect exciton density oscillations calculated for different temperature of the phonon bath. Parameters correspond to regime I.}
\label{Fig3}
\end{figure}

The time dependence of indirect exciton density oscillations $n_{IX}(t)$, calculated numerically from the Eqns. (\ref{aC})--(\ref{cIX}), can be fitted with the analytical function $n_{IX}(t)=n_{IX}^{0}\cos^{2}(\omega t/2)e^{-t/\tau }$,
%
%
where $\omega =2\pi \nu \approx \sqrt{J^2 + \delta_{J}^2(F)}$ is the frequency of oscillations, $n_{IX}^{0}$ denotes the magnitude of oscillations which decreases in time with the damping rate $\tau^{-1}$. Tuning the electric field, the frequency of generated oscillations changes in the range of several THz due to its dependence on IX-DX detuning [Fig. \ref{Fig3}(a)], given as $\delta_{J}(F)= \delta_{I-D}(1-F/F_{0})$ \cite{ButovJPCM}. Other important parameters of the system such as the amplitude of oscillations and dimensionless oscillation quality factor $\xi$ defined as a ratio of magnitude to the decrement of oscillations can be tuned by variation of pumping conditions, as well as by the applied electric field \cite{SM}.

Rising the temperature of the system, the phonon scattering and subsequent population of the reservoir states lead to increase of characteristic decay rate of the oscillations [Fig. \ref{Fig3}(b)]. However, the beats of excitons density are still observable for comparably high frequency, and can be exploited in the pulsed regime.

Reducing the detuning between exciton and photon modes $\delta_{\Omega}$ one can bring the system into the regime II. In this regime, the IX density oscillations are observed as well, while their quality factor is different. Strong interactions of the cavity mode with the IX-DX resonance result in a higher magnitude of $n_{IX}$ oscillations than in the regime I [Fig. \ref{Fig2}(b)]. The damping rate of these oscillations is also larger. Clearly, the regime of strong coupling between all modes is advantageous for the pulsed pumping regime and it allows for high power generation. On the other hand, the regime I is preferential for the long-standing signal generation providing higher quality factor $\xi$, while lower amplitude of the emission.

We have shown that due to the coupling between modes IX and DX densities oscillate with a THz frequency. This infers that a dipolariton is an oscillating dipole, with a dipole moment in $z$ direction $d_{z}$ changing periodically from value $d_{z}=d_{0}$ of IX to $d_{z}=0$ corresponding to DX. One can define the total dipole moment of the system as $D_{z}=N_{IX}d_{z}$, where $N_{IX}=n_{IX}A$ is the number of indirect excitons within the area $A$ illuminated by the pumping light. Since we have shown before that after initial transient regime the density of indirect excitons $n_{IX}(t)$ is a decaying harmonic function of time, the total dipole moment of the system can be found as $D_{z}(t)=d_{0}n_{IX}A\cos ^{2}(\omega t/2)e^{-t/\tau}$.
%
%
Here $n_{IX}$ is the maximum density of indirect excitons.

The total intensity of the far field radiation emitted by a classical Hertz dipole can be found as $I=\ddot{D}_{z}^{2}/6\pi \epsilon_{0}c^{3}$ \cite{Landau},
%
%
where $D_{z}$ is the total dipole moment of the dipolariton array, $\epsilon_{0}$ is vacuum permittivity and $c$ is the speed of light. For the particular case of an array of harmonic dipole oscillations the intensity is $I=N_{IX}^{2}d_{0}^{2}\omega^{4}/48\pi \epsilon_{0}c^{3}$,
%
%
where $d_{0}=eL$ is a dipole moment of indirect exciton. Here for simplicity we did not consider the damping part $e^{-t/\tau }$ of oscillations. It will result in damping of the total power of emission, which is why to achieve stable cw radiation one therefore needs to using a sequence of pump pulses. Similarly to the conventional case of an elementary dipole emitter, the polar pattern is given by $I_{\theta }\sim \sin^{2}\theta $, where $\theta $ is an angle between direction of radiation and the growth axis of the structure \cite{SM}.

It is important to note that the total emitted intensity is proportional to the square of the density of indirect excitons and is dependent on the pump intensity in a non-linear way. This is a manifestation of the \emph{superradiance} \cite{Dicke,Bohnet,Yukalov} effect: due to the interference of coherent in-phase oscillations of elementary dipoles the output power is enhanced. This effect is sensitive to the quality factor of the cavity: the longer Rabi oscillations persist the stronger the amplification effect is. Superradiance is a specific feature of the dipolariton THz emitter, which makes it more efficient than any existing laser-to-THz converter.

For $\hbar J=6$ meV and the typical distance between QWs of $L=14$ nm, the power of THz emission of an elementary dipole formed by a dipolariton is $I_{0}=1.5\times 10^{-18}$ W $=1.5$ aW.  The total power emitted is given by this quantity multiplied by the square number of elementary dipoles $N_{IX}$. For the typical concentration of indirect excitons achievable experimentally of $n_{IX}=10^{10}$ $cm^{-2}$ and a $68~\mu$m diameter of the pumping spot one obtains $I_{tot}\approx 2~\mu$W. By embedding a stack of double quantum wells $n_{DQW}=4$ in each microcavity and using several sets of cavities on one chip one can obtain the output power similar to one of quantum cascade lasers. Furthermore, to improve the efficiency of THz radiation, the system can be placed in an external cavity tuned to the THz frequency \cite{Geiser,Walther}, with efficiency of emission being increased by the Purcell factor of the external cavity (see \cite{SM}, Fig. S5, for the sketch of setup).

\emph{Conclusions.---}We have shown that the system of dipolaritons can provide an efficient tunable source of THz radiation. In this system, due to the direct-indirect exciton coupling, the optically excited dipolariton system exhibits density oscillations with a subpicosecond period. The dipole oscillations lead to the superradiant emission of radiation in the THz frequency range. The spectral properties as well as the power output of this system are expected to be strongly improved with respect to the existing laser induced THz emitters. Additionally, we propose a way to enhance the radiation efficiency using a supplementary THz cavity in the weak coupling regime.

We thank Jeremy J. Baumberg and Sven H\"{o}fling for useful discussions on the subject. This work has been supported by FP7 IRSES projects ``POLATER'' and ``POLAPHEN''. O. K. acknowledges the support from Eimskip Fund. A. K. acknowledges support from Russian ministry of Science and Education. I. A. S. acknowledge the support of Tier1 project ``Novel polaritonic devices''.

\newpage
\renewcommand{\theequation}{S\arabic{equation}}
\renewcommand\thefigure{S\arabic{figure}}

\begin{widetext}
\begin{center}
\textbf{Supplemental Material: Superradiant terahertz emission by dipolaritons}
\end{center}

\appendix

\subsection{A: Derivation of dynamic equations (lowest order of mean-field theory)}
Here we present the derivation of Eqns. (4)--(8) of the main text, which describe the dynamics of dipolariton system in the lowest order of mean field theory. 

\textit{Coherent part.} In order to find the macroscopic occupation of the modes one can treat fields as classical and calculate the dynamics of the system using mean values of their creation (or annihilation) operators. Using the Hamiltonian (1) and the master equation, one can write the equation of motion for the mean value of operator $\hat{a}$ as
\begin{equation}
\hbar \partial_{t} \langle \hat{a} \rangle(t)=i{\rm Tr} \lbrace \hat{a} [\rho, \hat{\cal{H}}_{coh}] \rbrace =i{\rm Tr} \lbrace \rho [\hat{\cal{H}}_{coh},\hat{a}] \rbrace,
\label{first}
\end{equation}
where we used the cyclic properties of the trace. The commutator in the RHS may be written using explicit form of the Hamiltonian as
\begin{equation}
\left[\omega_{C}\hat{a}^{\dagger}\hat{a}+ \omega_{DX}\hat{b}^{\dagger}\hat{b}+ \omega_{IX}\hat{c}^{\dagger}\hat{c}+\frac{\Omega}{2}(\hat{a}^{\dagger}\hat{b}+\hat{b}^{\dagger}\hat{a})-\frac{J}{2}(\hat{b}^{\dagger}\hat{c}+\hat{c}^{\dagger}\hat{b})+\widetilde{P}\hat{a}^{\dagger}+\widetilde{P}^{*}\hat{a},\hat{a} \right],
\end{equation}
and it can be calculated part by part. Naturally, only the first, fourth and penultimate terms contribute to the equation. Using standard commutation relations for bosons, one obtains $\omega_{C}[\hat{a}^{\dagger}\hat{a}, \hat{a}]=-\omega_{C}\hat{a}$, $\frac{\Omega}{2}[\hat{a}^{\dagger}\hat{b}, \hat{a}]=-\frac{\Omega}{2} \hat{b}$, and $\widetilde{P}[\hat{a}^{\dagger},\hat{a}]=-\widetilde{P}$. The coherent part of the equations for $\langle \hat{b} \rangle(t)$ and $\langle \hat{c} \rangle(t)$ can be written in the same way.

\textit{Dissipation.} The incoherent processes can be treated using Lindblad superoperator \cite{Carmichael}. In particular, the decay of the cavity mode can be included in the Eq. (\ref{first}) as
\begin{equation}
\partial_{t} \langle \hat{a} \rangle(t)={\rm Tr}\lbrace \hat{a} \frac{\partial \rho(t)}{\partial t} \rbrace = {\rm Tr}\lbrace \hat{a} \frac{\gamma_{C}}{2}(2\hat{a}\rho \hat{a}^{\dagger}-\hat{a}^{\dagger}\hat{a}\rho -\rho \hat{a}^{\dagger}\hat{a}) \rbrace.
\end{equation}
The first term gives $\gamma_{C}{\rm Tr}\lbrace \rho \hat{a}^{\dagger}\hat{a}\hat{a} \rbrace = \gamma_{C}{\rm Tr}\lbrace \rho \hat{a}\hat{a}^{\dagger}\hat{a} \rbrace - \gamma_{C}{\rm Tr}\lbrace \rho \hat{a} \rbrace$, where we used the relation $(\hat{a}^{\dagger}\hat{a})\hat{a}=\hat{a}(\hat{a}^{\dagger}\hat{a}-1)$. The second term reads as $-\frac{\gamma_{C}}{2} {\rm Tr}\lbrace \rho \hat{a}\hat{a}^{\dagger}\hat{a} \rbrace$, and third term yields $-\frac{\gamma_{C}}{2} {\rm Tr}\lbrace \rho \hat{a}^{\dagger}\hat{a}\hat{a} \rbrace = -\frac{\gamma_{C}}{2} {\rm Tr}\lbrace \rho \hat{a}\hat{a}^{\dagger}\hat{a} \rbrace + \frac{\gamma_{C}}{2} {\rm Tr}\lbrace \rho \hat{a} \rbrace$. The dissipative part of dynamics is described by a kinetic equation
\begin{equation}
\partial_{t} \langle \hat{a} \rangle(t)=-\frac{\gamma_{C}}{2} \langle \hat{a} \rangle.
\end{equation}
The lifetime of $\langle \hat{b} \rangle(t)$ and $\langle \hat{c} \rangle(t)$ modes can be introduced in the same fashion.

\textit{Interaction with phonons.} Next, we want to account for interaction of indirect and direct excitons with a thermal reservoir of acoustic phonons in the system. In general, it requires the accounting of full momentum space of exciton-photon system and represents a formidable task for a two-dimensional system. In the following we shall assume pumping at a normal incidence angle, so that the optically generated excitons are having zero in-plane momentum. The emission or absorption of acoustic phonon therefore leads to transition of particles to the thermalized reservoir states.

The generic Hamiltonian of exciton-phonon interaction can be written as
\begin{align}
\notag
\hat{\cal{H}}_{th}(t)= D_{DX} \sum_{k} (\hat{b}_{R,k}^{\dagger}\hat{b}+ \hat{b}^{\dagger}\hat{b}_{R,k}) e^{i(E_{DX,k}-E_{DX})t} (\hat{d}_{k} e^{-i \omega_{k}t} + \hat{d}^{\dagger}_{k} e^{i \omega_{k}t}) \\
 + D_{IX} \sum_{k} (\hat{c}^{\dagger}\hat{c}_{R,k} + \hat{c}_{R,k}^{\dagger} \hat{c}) e^{i(E_{IX,k}-E_{IX})t} (\hat{d}_{k} e^{-i \omega_{k}t} + \hat{d}^{\dagger}_{k} e^{i \omega_{k}t}),
\label{HthG}
\end{align}
where $\hat{d}^{\dagger}_{k}$ and $\hat{d}_{k}$ are creation and annihilation operators for acoustic phonon with momentum $k$. Here $D_{DX,IX}$ are direct (indirect) exciton-phonon interaction constants, and $E_{DX,IX}$ are energies of direct and indirect excitons at $k=0$. Additionally, we introduced operators for particles in the reservoir, $\hat{b}^{\dagger}_{R,k}$ and $\hat{c}^{\dagger}_{R,k}$, with energies given by $E_{DX,k}$ and $E_{IX,k}$.

Processes of exciton-phonon scattering contribute to the incoherent part of the dynamics of the system and can be described using the Liouville-von Neumann equation written in an integro-differential form \cite{Savenko}
\begin{equation}
(\partial_{t}\rho)^{(th)}=-\frac{1}{\hbar^2}\int_{-\infty}^{t}dt'\left[\hat{\cal{H}}_{th}(t),\left[\hat{\cal{H}}_{th}(t'),\rho(t)\right] \right],
\label{LvN}
\end{equation}
where $\hat{\cal{H}}_{th}(t)$ corresponds to generic exciton-phonon interaction Hamiltonian and by index $^{(th)}$ we denote the part of dynamics coming from the thermal bath interaction. Applying Born-Markov approximation we obtain
\begin{equation}
(\partial_{t}\rho)^{(th)}=\delta_{\Delta E} \left[ 2 (\hat{\cal{H}}^{+} \rho \hat{\cal{H}}^{-} + \hat{\cal{H}}^{-} \rho \hat{\cal{H}}^{+})- (\hat{\cal{H}}^{+} \hat{\cal{H}}^{-}+ \hat{\cal{H}}^{-}\hat{\cal{H}}^{+})\rho - \rho (\hat{\cal{H}}^{+} \hat{\cal{H}}^{-} + \hat{\cal{H}}^{-} \hat{\cal{H}}^{+}) \right],
\end{equation}
where $\delta_{\Delta E}$ is a constant accounting for the energy conservation and is taken to be inverse broadening of exciton states divided by square of Plank constant. One can see that the time-dependence of the Hamiltonian disappears because of the energy conservation. Here $\hat{\cal{H}}^{+}$ and $\hat{\cal{H}}^{+}$ denote exciton-phonon interaction Hamiltonians corresponding emission and absorption of phonon, which can be written as $\hat{\cal{H}}_{th}=\hat{\cal{H}}^{+} + \hat{\cal{H}}^{-}$, and 
\begin{align}
\label{H+}
&\hat{\cal{H}}^{+} = D_{DX} \sum_{k} \hat{b}^{\dagger}\hat{b}_{R,k} \hat{d}^{\dagger}_{k} + D_{IX} \sum_{k} \hat{c}^{\dagger}\hat{c}_{R,k} \hat{d}^{\dagger}_{k},\\
&\hat{\cal{H}}^{-} =  D_{DX} \sum_{k} \hat{b}_{R,k}^{\dagger}\hat{b} \hat{d}_{k} + D_{IX} \sum_{k} \hat{c}_{R,k}^{\dagger}\hat{c} \hat{d}_{k}.
\label{H-}
\end{align}
In the following Hamiltonian we accounted only for the energy conserving terms.

Let us now derive dynamic equation with additional treatment of temperature effects and reservoir states. The corresponding contribution to the time evolution of the mean value of any operator $\langle \hat{\cal{O}} \rangle = \rm{Tr}{\rho \hat{\cal{O}}}$ reads as
\begin{equation}
(\partial_{t} \langle \hat{\cal{O}} \rangle)^{(th)}= \delta_{\Delta E} \left( \langle \left[ \hat{\cal{H}}^{-},\left[\hat{\cal{O}}, \hat{\cal{H}}^{+} \right] \right] \rangle + \langle \left[ \hat{\cal{H}}^{+},\left[\hat{\cal{O}}, \hat{\cal{H}}^{-} \right] \right] \rangle \right).
\end{equation}
We proceed with calculation of phonon-assisted part of dynamics for DX and IX modes. For instance, the phonon contribution to dynamics of mean value of IX field $\langle \hat{\cal{O}} \rangle= \langle \hat{c} \rangle$ yields
\begin{equation}
(\partial_{t} \langle \hat{c} \rangle)^{(th)}= \delta_{\Delta E} \left( \langle \left[ \hat{\cal{H}}^{-},\left[\hat{c}, \hat{\cal{H}}^{+} \right] \right] \rangle + \langle \left[ \hat{\cal{H}}^{+},\left[\hat{c}, \hat{\cal{H}}^{-} \right] \right] \rangle \right).
\label{cth}
\end{equation}
The calculation of commutators is straightforward and gives $[\hat{\cal{H}}^{+},[\hat{b},\hat{\cal{H}}^{-}]]=0$, $[\hat{\cal{H}}^{-},[\hat{b},\hat{\cal{H}}^{+}]]=D_{IX}^2 \sum_{k} (\hat{c}_{R,k}^{\dagger}\hat{c}_{R,k} - \hat{n}_{ph,k})\hat{c}$, where $\hat{n}_{ph,k}=\hat{d}_{k}^{\dagger}\hat{d}_{k}$ and we account only for diagonal terms of phonon density matrix. Similar expression can be found for direct exciton field dynamics.

In order to close the set of equations we need to derive the expression for dynamical equations of reservoirs. For instance, the DX and IX reservoir states equations read as
\begin{align}
\label{bRth}
(\partial_{t} \langle \hat{b}_{R,q}^{\dagger} \hat{b}_{R,q} \rangle)^{(th)}= &\delta_{\Delta E} \left( \langle \left[ \hat{\cal{H}}^{-},\left[\hat{b}_{R,k}^{\dagger} \hat{b}_{R,k}, \hat{\cal{H}}^{+} \right] \right] \rangle + \langle \left[ \hat{\cal{H}}^{+},\left[\hat{b}_{R,k}^{\dagger} \hat{b}_{R,k}, \hat{\cal{H}}^{-} \right] \right] \rangle \right) \approx \\
\notag
&- \delta_{\Delta E} D_{DX}^2 \sum_{k} \lbrace |\langle \hat{b} \rangle|^2 (\langle \hat{b}_{R,k}^{\dagger} \hat{b}_{R,k} \rangle +1) n_{\text{ph},k} - (|\langle \hat{b} \rangle|^2 +1) \langle \hat{b}_{R,k}^{\dagger} \hat{b}_{R,k} \rangle (n_{\text{ph},k}+1) \rbrace,\\
\label{cRth}
(\partial_{t} \langle \hat{c}_{R,q}^{\dagger} \hat{c}_{R,q} \rangle)^{(th)} \approx &- \delta_{\Delta E} D_{IX}^2 \sum_{k} \lbrace |\langle \hat{c} \rangle|^2 (\langle \hat{c}_{R,k}^{\dagger} \hat{c}_{R,k} \rangle +1) n_{\text{ph},k} - (|\langle \hat{c} \rangle|^2 +1) \langle \hat{c}_{R,k}^{\dagger} \hat{c}_{R,k} \rangle (n_{\text{ph},k}+1) \rbrace.
\end{align}

Finally, merging all contribution to dynamics of the fields and reservoir states, we come to the system of five coupled equations (4)--(8) in the main text.


\subsection{B: The full model accounting for phonons and dephasing}
In the previous section we derived equations which describe the dipolariton system in the lowest order of the mean-field theory. Here we generalize it to the next order and derive equation for the mean values of occupation numbers of the modes and their correlators.

\textit{Coherent part.} We start from the Hamiltonian given by Eq. (1). The time dependence of the occupation numbers and correlators can be found using relation
\begin{equation}
\partial_{t} n_{i}(t)={\rm Tr} \lbrace \hat{a}^{\dagger}_{i}\hat{a}_{i} \partial_{t}\rho(t) \rbrace,
\end{equation}
where $\hat{a}_{i}$ represents operator of different type ($i=C,DX,IX$).
The equation can be rewritten in form
\begin{equation}
\partial_{t} n_{i}(t)=\frac{i}{\hbar}{\rm Tr} \lbrace \hat{a}^{\dagger}_{i}\hat{a}_{i} [\rho ,\hat{\cal{H}}_{coh}] \rbrace,
\end{equation}
and simplified using the explicit expression for coherent part of the Hamiltonian. To describe the dynamics we need to write a system of six coupled equations for occupation numbers of the modes $n_{C}=\langle \hat{a}^{\dagger}\hat{a}\rangle $, $n_{DX}=\langle \hat{b}^{\dagger} \hat{b} \rangle$, $n_{IX}=\langle \hat{c}^{\dagger} \hat{c} \rangle$ and correlators $\alpha_{ab}=\langle \hat{a}^{\dagger}\hat{b}\rangle$, $\alpha_{bc}=\langle \hat{b}^{\dagger}\hat{c}\rangle$ and $\alpha_{ac}=\langle \hat{a}^{\dagger}\hat{c}\rangle$. Using the mean-field approximation, we derive equation for each type of occupation number and correlator. For instance, the occupation number of indirect excitons can be written as
\begin{equation}
\partial_{t} n_{IX}(t)=i{\rm Tr} \lbrace \rho \left[\omega_{C}\hat{a}^{\dagger}\hat{a}+ \omega_{DX}\hat{b}^{\dagger}\hat{b}+ \omega_{IX}\hat{c}^{\dagger}\hat{c}+\frac{\Omega}{2}(\hat{a}^{\dagger}\hat{b}+\hat{b}^{\dagger}\hat{a})-\frac{J}{2}(\hat{b}^{\dagger}\hat{c}+\hat{c}^{\dagger}\hat{b})+ \widetilde{P}\hat{a}^{\dagger}+\widetilde{P}^{*}\hat{a},\hat{c}^{\dagger}\hat{c} \right] \rbrace.
\end{equation}
The first four terms of commutator are zero, while fifth term gives
\begin{equation}
-\frac{J}{2}\left( \left[\hat{b}^{\dagger}\hat{c},\hat{c}^{\dagger}\hat{c} \right]+\left[\hat{c}^{\dagger}\hat{b},\hat{c}^{\dagger}\hat{c} \right] \right)=-\frac{J}{2}\left( [\hat{b}^{\dagger}\hat{c},\hat{c}^{\dagger}]\hat{c}+ \hat{c}^{\dagger}[\hat{b}^{\dagger}\hat{c},\hat{c}]+ [\hat{c}^{\dagger}\hat{b},\hat{c}^{\dagger}]\hat{c}+\hat{c}^{\dagger}[\hat{c}^{\dagger}\hat{b},\hat{c}] \right)=-\frac{J}{2}\left( \hat{b}^{\dagger}\hat{c}-\hat{c}^{\dagger}\hat{b} \right),
\end{equation}
with final equation being
\begin{equation}
\partial_{t} n_{IX}(t)=J {\rm Im}\alpha_{bc},
\end{equation}
where we used relation ${\rm Im}A=(A-A^{*})/2i$. One can write equations for $n_{C}$ and $n_{DX}$ in the same fashion.

Derivation of dynamic equations for correlators require more efforts, but can be done straightforwardly. For example, the equation for correlator $\alpha_{ab}=\langle \hat{a}^{\dagger}\hat{b} \rangle$ reads as 
\begin{equation}
\partial_{t} \alpha_{ab}(t)=i{\rm Tr} \lbrace \rho \left[\omega_{C}\hat{a}^{\dagger}\hat{a}+ \omega_{DX}\hat{b}^{\dagger}\hat{b}+ \omega_{IX}\hat{c}^{\dagger}\hat{c}+\frac{\Omega}{2}\hat{a}^{\dagger}\hat{b}+ \frac{\Omega}{2}\hat{b}^{\dagger}\hat{a}- \frac{J}{2}\hat{b}^{\dagger}\hat{c}+ \frac{J}{2}\hat{c}^{\dagger}\hat{b}+ \widetilde{P}\hat{a}^{\dagger}+\widetilde{P}^{*}\hat{a},\hat{a}^{\dagger}\hat{b} \right] \rbrace.
\label{S_alpha_ab}
\end{equation}
Considering terms one by one, the first gives $\omega_{C}[\hat{a}^{\dagger}\hat{a},\hat{a}^{\dagger}\hat{b}]=\omega_{C}\hat{b}\hat{a}^{\dagger}$ and second term yields $\omega_{DX}[\hat{b}^{\dagger}\hat{b},\hat{a}^{\dagger}\hat{b}]=-\omega_{DX}\hat{a}^{\dagger}\hat{b}$. Naturally, third and forth terms give zero, while the fifth term is equal to $\Omega/2[\hat{a}\hat{b}^{\dagger},\hat{a}^{\dagger}\hat{b}]=-\Omega/2(\hat{a}^{\dagger}\hat{a}-\hat{b}^{\dagger}\hat{b})$. The sixth term reads $-J/2[\hat{b}^{\dagger}\hat{c},\hat{a}^{\dagger}\hat{b}]=J/2~\hat{a}^{\dagger}\hat{c}$, and seventh term from correlator in Eq. (\ref{S_alpha_ab}) is zero. This leads to dynamic equation for the correlator
\begin{equation}
\partial_t \alpha_{ab}= -i\delta_{\Omega} \alpha_{ab}-i\frac{\Omega}{2}(n_{C}-n_{DX}) -i\frac{J}{2} \alpha_{ac}+ i  \widetilde{P}^{*}\langle \hat{b} \rangle.
\end{equation}
Equations for $\partial_t \alpha_{bc}$ and $\partial_t \alpha_{ac}$ can be derived analogously. 

Finally, assuming the presence of coherent pump, we need to add three equation for $\langle \hat{a} \rangle$, $\langle \hat{b} \rangle$ and $\langle \hat{c} \rangle$ given by Eqns. (4)--(6) in the main text (excluding thermal part). This leads to complete system of Eqns. (\ref{nC})--(\ref{cIX}) for dynamics of the dipolariton system written at the end of the section.

\textit{Decoherent part.} The master equation for density matrix is a powerful tool which allows one to account for various incoherent processes. They can be introduced in the model using Lindblad superoperator formalism \cite{Savenko,Kavokin}. 
First, let us account for dissipation processes coming from the leakage of cavity mode out of the mirrors, as well as radiative and non-radiative recombination of excitons. 

\emph{Dissipation processes} can be modelled with the Lindblad superoperator written for any operator $\hat{a}_{i}$ of type $i$ in the form
\begin{equation}
\hat{\cal{L}}\rho^{(dis)} =i \frac{\Gamma_i}{2} \left( 2 \hat{a}\rho\hat{a}^{\dagger}-\hat{a}^{\dagger}\hat{a}\rho -\rho\hat{a}^{\dagger}\hat{a} \right),
\end{equation}
where $\Gamma_{i}=\hbar \gamma_{i}$ represent the decay rate of the mode, $i=C,DX,IX$. Reminding the master equation, $i\hbar \partial \rho/\partial t=[\hat{\cal{H}},\rho]+\hat{\cal{L}}\rho$, one can write the incoherent part of dynamic equation for occupation numbers and correlators. For instance, in the case of cavity photon occupation number it reads
\begin{equation}
\partial_{t} n_{C}(t)={\rm Tr}\lbrace \hat{a}^{\dagger}\hat{a} \frac{\partial \rho(t)}{\partial t} \rbrace =
{\rm Tr}\lbrace \hat{a}^{\dagger}\hat{a} \frac{\gamma_{C}}{2}(2\hat{a}\rho \hat{a}^{\dagger}-\hat{a}^{\dagger}\hat{a}\rho -\rho \hat{a}^{\dagger}\hat{a}) \rbrace = \gamma_{C}{\rm Tr}\lbrace \rho (\hat{a}^{\dagger}\hat{a}^{\dagger}\hat{a}\hat{a}-\hat{a}^{\dagger}\hat{a}\hat{a}^{\dagger}\hat{a}) \rbrace ,
\end{equation}
where to write the latter expression we used the cyclic properties of the trace. Applying the mean-field approximation, $\langle \hat{n}_{C} \rangle=\langle \hat{a}^{\dagger}\hat{a} \rangle={\rm Tr}\lbrace \hat{n}_{C}\rho \rbrace$ and $\langle \hat{a}^{\dagger}\hat{a}^{\dagger}\hat{a}\hat{a} \rangle= {\rm Tr}\lbrace \hat{n}_{C}(\hat{n}_{C}-1)\rho \rbrace$, it can be rewritten as
\begin{equation}
\partial_{t} n_{C}(t)= \gamma_{C}{\rm Tr}\lbrace \rho (\hat{n}_{C}(\hat{n}_{C}-1)-\hat{n}_{C}^2) \rbrace =-\gamma_{C} n_{C}.
\end{equation}
The same considerations are valid for $\partial_{t} n_{DX}(t)$, $\partial_{t} n_{IX}(t)$ and correlators.

The radiative decay of the mode is not the only cause which affects the incoherent dynamics of the system. Indeed, the coherence in the system can be spoiled due to dephasing even for infinitely long-lived atomic systems. There the main effect of loosing coherence can be introduced by a \emph{pure decoherence} term, which acts solely on off-diagonal terms or ``coherences'', while live diagonal terms are unaffected. Usually written for two-level system (atom, qubit etc), the Lindblad operator for pure coherence in general form reads \cite{Schlosshauer}
\begin{equation}
\hat{\cal{L}}\rho^{(dec)} =-i \Gamma^{(dec)} \left[ \sigma_{z},[\sigma_{z},\rho] \right],
\label{Ldec}
\end{equation}
where $\sigma_{z}=[\sigma^{\dagger},\sigma]/2$, $\sigma^{\dagger}$ is a fermionic creation operator, and $\Gamma^{(dec)}$ denotes the rate of pure dephasing. In particular, for two-level system it thus can be rewritten as $\hat{\cal{L}}_{dec}\rho =i \frac{\Gamma^{(dec)}}{2} \left( \sigma_{z} \rho \sigma_{z} -\rho \right)$ \cite{Laussy}. In our system we want to account for the decay of coherence of the interaction between different modes which have bosonic statistics. Therefore, we can map the angular momentum operators into bosonic operators using Holstein-Primakoff transformations \cite{HP}, $\sigma_{z}=(s-\hat{a}^{\dagger}\hat{a})$, where $s=1$ is spin of the particle. Then Lindblad operator (\ref{Ldec}) can be rewritten opening the commutator and reads
\begin{equation}
\hat{\cal{L}}\rho^{(dec)} =i \Gamma^{(dec)} \left( 2\hat{a}^{\dagger}\hat{a}\rho \hat{a}^{\dagger}\hat{a}-\hat{a}^{\dagger}\hat{a}\hat{a}^{\dagger}\hat{a}\rho - \rho \hat{a}^{\dagger}\hat{a}\hat{a}^{\dagger}\hat{a} \right).
\label{Ldec2}
\end{equation}
Plugging (\ref{Ldec2}) into equation for occupation number
\begin{equation}
\partial_{t} n_{C}(t)= \gamma^{(dec)} {\rm Tr}\lbrace \hat{a}^{\dagger}\hat{a} \left( 2\hat{a}^{\dagger}\hat{a}\rho \hat{a}^{\dagger}\hat{a}-\hat{a}^{\dagger}\hat{a}\hat{a}^{\dagger}\hat{a}\rho - \rho \hat{a}^{\dagger}\hat{a}\hat{a}^{\dagger}\hat{a} \right) \rbrace =0
\end{equation}
one can check that it gives zero contribution to the lifetime of the mode. Here we used definition $\gamma^{(dec)}=\Gamma^{(dec)}/\hbar$. Performing the same analysis for correlator (e. g. $\alpha_{ab}$) we straightforwardly derive
\begin{equation}
\partial_{t} \alpha_{ab}(t)= \gamma^{(dec)}_{ab}{\rm Tr}\lbrace \hat{a}^{\dagger}\hat{b} \left( 2\hat{a}^{\dagger}\hat{a}\rho \hat{a}^{\dagger}\hat{a}-\hat{a}^{\dagger}\hat{a}\hat{a}^{\dagger}\hat{a}\rho - \rho \hat{a}^{\dagger}\hat{a}\hat{a}^{\dagger}\hat{a} \right) \rbrace = \gamma^{(dec)}_{ab}{\rm Tr} \lbrace \rho \hat{a}^{\dagger}\hat{b}(2\hat{a}\hat{a}^{\dagger}\hat{a}^{\dagger}\hat{a}- \hat{a}^{\dagger}\hat{a}\hat{a}^{\dagger}\hat{a}- \hat{a}\hat{a}^{\dagger}\hat{a}\hat{a}^{\dagger}) \rbrace = -\gamma^{(dec)}_{ab} \langle \hat{a}^{\dagger}\hat{b} \rangle.
\end{equation}
Thus, coherence between mode is destroyed for large damping rates $\gamma^{(dec)}_{ab}$, and index corresponds to the type of correlator we are interested in. Similarly, one can introduce pure decoherence for $\alpha_{bc}$ and $\alpha_{ac}$ correlators.


\textit{Thermal part.} Following the same procedure for accounting of exciton-phonon interaction, which we introduced in the previous section, we derive thermal part of dynamics for occupation numbers and correlators. For example, the thermal contribution to dynamics of indirect exciton occupation number $n_{IX}$ is given by
\begin{equation}
(\partial_{t} n_{DX})^{(th)}= \delta_{\Delta E} \left( \langle \left[ \hat{\cal{H}}^{-},\left[\hat{b}^{\dagger}\hat{b}, \hat{\cal{H}}^{+} \right] \right] \rangle + \langle \left[ \hat{\cal{H}}^{+},\left[\hat{b}^{\dagger}\hat{b}, \hat{\cal{H}}^{-} \right] \right] \rangle \right),
\label{nDXth}
\end{equation}
and commutators inside can be calculated as
\begin{equation}
\left[ \hat{b}^{\dagger} \hat{b}, \hat{\cal{H}}^{+} \right] = \left[ \hat{b}^{\dagger} \hat{b}, D_{DX} \sum_{k} \hat{b}^{\dagger}\hat{b}_{R,k} \hat{d}^{\dagger}_{k} + D_{IX} \sum_{k} \hat{c}^{\dagger}\hat{c}_{R,k} \hat{d}^{\dagger}_{k} \right] = D_{DX} \sum_{k} \hat{b}^{\dagger} \hat{b}_{R,k} \hat{d}^{\dagger}_{k}
\end{equation}
and
\begin{align}
&\left[ \hat{\cal{H}}^{-}, \left[ \hat{b}^{\dagger} \hat{b}, \hat{\cal{H}}^{+} \right] \right] = \left[D_{DX} \sum_{k} \hat{b}_{R,k}^{\dagger}\hat{b} \hat{d}_{k} + D_{IX} \sum_{k} \hat{c}_{R,k}^{\dagger}\hat{c} \hat{d}_{k} , D_{DX} \sum_{k} \hat{b}^{\dagger} \hat{b}_{R,k} \hat{d}^{\dagger}_{k} \right] \\ \notag
&=D_{DX}^{2} \sum_{k} \lbrace (\hat{b}^{\dagger} \hat{b} +1)\hat{b}_{R,k}^{\dagger}\hat{b}_{R,k} (\hat{d}_{k}^{\dagger}\hat{d}_{k} +1) - \hat{b}^{\dagger}\hat{b} (\hat{b}_{R,k}^{\dagger}\hat{b}_{R,k}+1)\hat{d}_{k}^{\dagger}\hat{d}_{k} \rbrace + D_{DX}D_{IX} \sum_{k} \hat{b}^{\dagger} \hat{c} \hat{c}^{\dagger}_{R,k} \hat{b}_{R,k},
\end{align}
where the last term corresponds to DX-IX phonon-mediated interaction involving reservoir states. 

The correlator $ [ \hat{\cal{H}}^{+},\left[\hat{b}^{\dagger}\hat{b}, \hat{\cal{H}}^{-} \right] ]$ yields the same contribution. Finally, applying trace operation we get the thermal part of the dynamic equation for the mean value of the operator $\hat{n}_{DX}$:
\begin{align}\notag
(\partial_{t} n_{DX})^{(th)}= &2 \delta_{\Delta E} D_{DX}^2 \sum_{k} \lbrace (n_{DX}+1)n^{DX}_{R,k}(n_{ph,k}+1) - n_{DX} (n^{DX}_{R,k}+1)n_{ph,k} \rbrace \\
&+ 2 \delta_{\Delta E} D_{DX} D_{IX} \sum_{k} \rm{Re} \lbrace \alpha_{bc} (\alpha_{bc,k}^{R})^{*} \rbrace,
\end{align}
where $n^{DX}_{R,k}=\langle \hat{b}_{R,k}^{\dagger} \hat{b}_{R,k} \rangle$ corresponds to number of particles in the reservoir and $n_{ph,k}$ denotes occupancy of phonon mode given by Bose-Einstein distribution function. Here we have also introduced a correlator between reservoir states $\alpha_{bc,k}^{R} = \langle \hat{b}^{\dagger}_{R,k} \hat{c}_{R,k} \rangle$. One sees that interaction with phonons leads to decrease of population of coherent DX and IX modes.

An analogous procedure can be performed with $n_{IX}$ occupation number, 
\begin{align}\notag
(\partial_{t} n_{IX})^{(th)}= &2 \delta_{\Delta E} D_{IX}^2 \sum_{k} \lbrace (n_{IX}+1)n^{IX}_{R,k}(n_{ph,k}+1) - n_{IX} (n^{IX}_{R,k}+1)n_{ph,k} \rbrace \\
&+ 2 \delta_{\Delta E} D_{DX} D_{IX} \sum_{k} \rm{Re} \lbrace \alpha_{bc} (\alpha_{bc,k}^{R})^{*} \rbrace,
\end{align}
while thermal contribution to cavity photon occupation number dynamics yields zero, $(\partial_{t} n_{C})^{(th)}=0$.

Let us now calculate the incoherent part of the dynamics of averages of operators $\langle \hat{b} \rangle$ and $\langle \hat{c} \rangle$. Starting with direct exciton annihilation operator we obtain
\begin{equation}
(\partial_{t} \langle \hat{b} \rangle)^{(th)}= \delta_{\Delta E} \left( \langle \left[ \hat{\cal{H}}^{-},\left[\hat{b}, \hat{\cal{H}}^{+} \right] \right] \rangle + \langle \left[ \hat{\cal{H}}^{+},\left[\hat{b}, \hat{\cal{H}}^{-} \right] \right] \rangle \right),
\end{equation}
where the second term gives zero, and $\left[ \hat{\cal{H}}^{-},\left[\hat{b}, \hat{\cal{H}}^{+} \right] \right] = D_{DX}^2 \sum_{k} \hat{b} (\hat{n}^{DX}_{R,k} - \hat{n}_{ph,k})$. The dynamical equations for mean values therefore read
\begin{align}
&(\partial_{t} \langle \hat{b} \rangle)^{(th)}= \delta_{\Delta E} D_{DX}^2 \sum_{k} \langle \hat{b} \rangle (n^{DX}_{R,k} - n_{ph,k}) \\
&(\partial_{t} \langle \hat{c} \rangle)^{(th)}= \delta_{\Delta E} D_{IX}^2 \sum_{k} \langle \hat{c} \rangle (n^{IX}_{R,k} - n_{ph,k}),
\end{align}
where $n^{IX}_{R,k} = \langle \hat{c}^{\dagger}_{R,k} \hat{c}_{R,k} \rangle$.

Next, we derive the thermal part of dynamics for correlators $\alpha_{ab}$, $\alpha_{ac}$ and $\alpha_{bc}$. The derivation of first two cases is trivial since cavity photon creation operator commutes with the phonon part of the Hamiltonian and yields
\begin{align}
&(\partial_{t} \alpha_{ab})^{(th)}= \delta_{\Delta E} D_{DX}^2 \sum_{k} \alpha_{ab} (n^{DX}_{R,k} - n_{ph,k}) \\
&(\partial_{t} \alpha_{ac})^{(th)}= \delta_{\Delta E} D_{IX}^2 \sum_{k} \alpha_{ac} (n^{IX}_{R,k} - n_{ph,k}).
\end{align}
The equation for correlator $\alpha_{bc}$ is
\begin{equation}
(\partial_{t} \alpha_{bc})^{(th)}= \delta_{\Delta E} \left( \langle \left[ \hat{\cal{H}}^{-},\left[\hat{b}^{\dagger}\hat{c}, \hat{\cal{H}}^{+} \right] \right] \rangle + \langle \left[ \hat{\cal{H}}^{+},\left[\hat{b}^{\dagger}\hat{c}, \hat{\cal{H}}^{-} \right] \right] \rangle \right),
\end{equation}
which after a straightforward algebra gives
\begin{align} \notag
(\partial_{t} \alpha_{bc})^{(th)}= &\delta_{\Delta E} D_{DX}^2 \sum_{k} \alpha_{bc}(n^{DX}_{R,k}-n_{ph,k}) + \delta_{\Delta E} D_{IX}^2 \sum_{k} \alpha_{bc}(n^{IX}_{R,k}-n_{ph,k}) \\ &+ 2 \delta_{\Delta E} D_{DX} D_{IX} \sum_{k} \lbrace \alpha^{R}_{bc,k} (n_{ph,k}+1)) \rbrace.
\end{align}
Consequently, rising the temperature of the system leads to the decay of correlations between modes.

Additionally, we need to supplement our system with an equation for reservoir occupation number $n^{DX}_{R,k}=\langle \hat{b}_{R,k}^{\dagger} \hat{b}_{R,k} \rangle$, $n^{IX}_{R,k}=\langle \hat{c}_{R,k}^{\dagger} \hat{c}_{R,k} \rangle$ and correlator $\alpha^{R}_{bc,k} = \langle \hat{b}^{\dagger}_{R,k} \hat{c}_{R,k} \rangle$. Operator corresponding to the reservoir states naturally commute with a coherent part of the Hamiltonian and are fully defined by exciton-phonon interaction Hamiltonian:
\begin{align} \notag
&\partial_{t} n^{DX}_{R,q} = -\gamma_{DX} n^{DX}_{R,q} + \delta_{\Delta E} \left( \langle \left[ \hat{\cal{H}}^{-},\left[\hat{b}_{R,k}^{\dagger}\hat{b}_{R,k}, \hat{\cal{H}}^{+} \right] \right] \rangle + \langle \left[ \hat{\cal{H}}^{+},\left[\hat{b}_{R,k}^{\dagger}\hat{b}_{R,k}, \hat{\cal{H}}^{-} \right] \right] \rangle \right) \approx -\gamma_{DX} n^{DX}_{R,q}  \\ \label{nDXRk}
&+ \delta_{\Delta E} D_{DX}^2 \sum_{k} \lbrace n_{DX} (n^{DX}_{R,k}+1) n_{ph,k}- (n_{DX}+1) n^{DX}_{R,k} (n_{ph,k}+1) \rbrace,\\
\label{nIXRk}
&\partial_{t} n^{IX}_{R,q} \approx -\gamma_{IX} n^{IX}_{R,q} + \delta_{\Delta E} D_{IX}^2 \sum_{k} \lbrace n_{IX} (n^{IX}_{R,k}+1) n_{ph,k}- (n_{IX}+1) n^{IX}_{R,k} (n_{ph,k}+1) \rbrace,\\
\label{alphaBCRk}
&\partial_{t} \alpha^{R}_{bc,q} = -\frac{\gamma_{bc}}{2} \alpha^{R}_{bc,q} + \delta_{\Delta E} D_{DX}^2 \sum_{k} \alpha^{R}_{bc,k} (n_{DX} + n_{ph,k} +1) - \delta_{\Delta E} D_{IX}^2 \sum_{k} \alpha^{R}_{bc,k} (n_{IX} + n_{ph,k} +1) \\
\notag & + \delta_{\Delta E} D_{DX} D_{IX} \sum_{k} \alpha_{bc}(n_{R,k}^{DX} - n_{ph,k}) - \delta_{\Delta E} D_{DX} D_{IX} \sum_{k} \alpha_{bc}(n_{R,k}^{IX} - n_{ph,k}),
\end{align}
where we introduced decay terms for reservoir states similarly to coherent exciton modes. One can see that there is almost no correlations between reservoir modes for equal exciton-phonon constants $D_{DX}\approx D_{IX}$.


Finally, merging coherent and decoherent parts, we obtain the system of equations
\begin{align}
\label{nC}
&\partial n_{C}/\partial t = \Omega {\rm Im} \alpha_{ab} - \gamma_{C} n_{C}-2{\rm Im}\lbrace \widetilde{P}^{*}\langle \hat{a} \rangle \rbrace,\\
\label{nDX}
&\partial n_{DX}/\partial t = -\Omega {\rm Im} \alpha_{ab} - J {\rm Im} \alpha_{bc} - \gamma_{DX} n_{DX} + (\partial_{t} n_{DX})^{(th)},\\
\label{IX}
&\partial n_{IX}/\partial t = J {\rm Im} \alpha_{bc} - \gamma n_{IX} + (\partial_{t} n_{IX})^{(th)},\\
\label{alpha_ab}
& \partial \alpha_{ab} /\partial t= -i\delta_{\Omega} \alpha_{ab}-i\frac{\Omega}{2}(n_{C}-n_{DX}) -i\frac{J}{2} \alpha_{ac}- \frac{\gamma_{ab}}{2}\alpha_{ab}- \gamma^{(dec)}_{ab} \alpha_{ab}+ i  \widetilde{P}^{*}\langle \hat{b} \rangle + (\partial_{t} \alpha_{ab})^{(th)} ,\\
\label{alpha_bc}
& \partial \alpha_{bc}/ \partial t= -i\delta_{J}\alpha_{bc} +i\frac{J}{2}(n_{IX}-n_{DX})+ i\frac{\Omega}{2} \alpha_{ac}- \frac{\gamma_{bc}}{2}\alpha_{bc}-\gamma^{(dec)}_{bc}\alpha_{bc} + (\partial_{t} \alpha_{bc})^{(th)},\\
\label{alpha_ac}
& \partial \alpha_{ac} /\partial t= i(\delta_{\Omega}-\delta_{J}) \alpha_{ac} -i\frac{J}{2} \alpha_{ab}+ i\frac{\Omega}{2} \alpha_{bc}-\frac{\gamma_{ac}}{2}\alpha_{ac}-\gamma^{(dec)}_{ac} \alpha_{ac}+ i\widetilde{P}^{*}\langle \hat{c} \rangle + (\partial_{t} \alpha_{ac})^{(th)},\\
\label{aC}
&\partial \langle \hat{a} \rangle /\partial t = -i \frac{\Omega}{2} \langle \hat{b} \rangle -\frac{\gamma_{C}}{2}\langle \hat{a} \rangle -i\widetilde{P},\\
\label{bDX}
&\partial \langle \hat{b} \rangle /\partial t = i\delta_{\Omega}\langle \hat{b} \rangle -i \frac{J}{2} \langle \hat{c} \rangle -i \frac{\Omega}{2} \langle \hat{a} \rangle -\frac{\gamma_{DX}}{2}\langle \hat{b} \rangle + (\partial_{t} \langle \hat{b} \rangle)^{(th)},\\
\label{cIX}
&\partial \langle \hat{c} \rangle /\partial t = i(\delta_{\Omega}-\delta_{J})\langle \hat{c} \rangle -i \frac{J}{2} \langle \hat{b} \rangle -\frac{\gamma_{IX}}{2}\langle \hat{c} \rangle + (\partial_{t} \langle \hat{c} \rangle)^{(th)},
\end{align}
and equations of motion for reservoir occupation numbers and correlator are given by Eqs. (\ref{nDXRk})-(\ref{alphaBCRk}).
Here $\gamma_{ab}=\gamma_{C}+\gamma_{DX}$, $\gamma_{bc}=\gamma_{DX}+\gamma_{IX}$ and $\gamma_{ac}=\gamma_{C}+\gamma_{IX}$ are reduced damping constants used in correlators, and we also introduced terms corresponding to pure decoherence terms with rates being $\gamma^{(dec)}_{ab}$, $\gamma^{(dec)}_{bc}$ and $\gamma^{(dec)}_{ac}$.
\begin{figure}[h]
\includegraphics[width=0.8\linewidth]{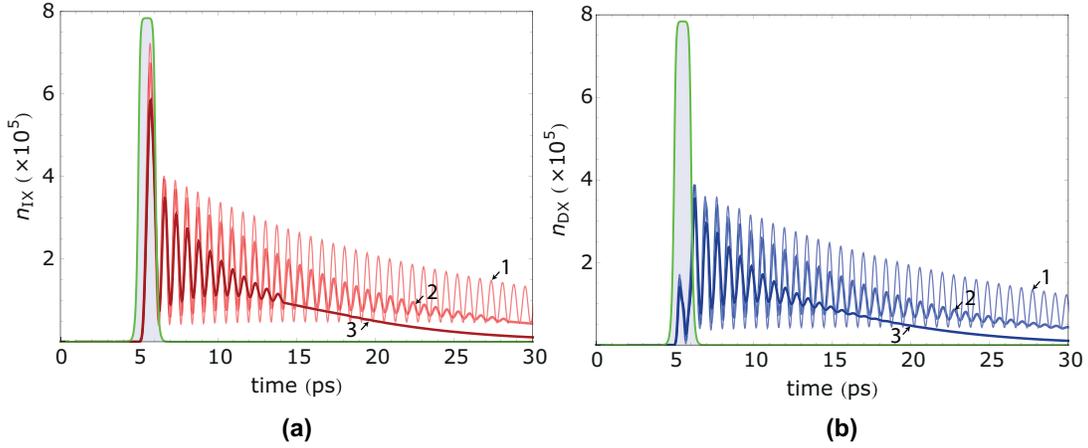}
\caption{(Color online) Dynamics of the dipolariton system subjected to pulsed optical pumping with accounting of temperature effects. Here we show time evolution of indirect exciton (a) and direct exciton (b) occupation numbers being $N_{DX}=n_{DX}A$ and $N_{IX}=n_{IX}A$, respectively, and $A$ denotes pumping area. They are calculated for temperatures $T=10$ K (curve 1), $T=70$ K (curve 2) and $T=140$ K (curve 3). Parameters of the system are chosen as for regime $I$ defined in the main text.}
\label{FigS1}
\end{figure}

We calculate the dynamics of the system accounting for non-zero temperature introduced by phonon distribution function. The result of simulation corresponding to regime $I$ is shown in Fig. \ref{FigS1} for indirect exciton (a) and direct exciton (b) occupation numbers. Here we considered exciton-phonon scattering rates being $W_{DX}=2 \delta_{\Delta E} D_{DX}^2 \approx$2 ps$^{-1}$, $W_{IX}=W_{DX}$, and pumping spot diameter is equal to 60 $\mu$m. 

While for low temperature ($T=10$ K, Fig. \ref{FigS1}, curve 1) influence of phonon interactions is minor and dynamics resembles previous zero temperature calculation, for higher temperatures we observe faster decay of exciton populations and damping of the IX-DX beats. We note that in the second-order mean field theory the exciton correlator $\alpha_{bc}$ experiences temperature enhanced dephasing. This, unlike in Fig. 3(b) of main text, leads both to decrease of amplitude and destruction of coherence of oscillations. Therefore, operation of dipolariton THz emitter for temperatures higher than hundred Kelvins is restricted to pulsed emission, similarly to regime $II$ described in the text.

Additionally, we should note that the numerical calculations for dynamics of the modes $n_{IX}(t)$ and $n_{DX}(t)$ calculated using system of Eqns. (4)--(8) written in the main text and advanced system (\ref{nC})--(\ref{cIX}) presented above give exactly the same result for zero temperature and chosen pumping conditions. Thus, this ensures the validity of used mean field approximation.

\subsection{C: Accounting for exciton-exciton interaction effects}
In the following section we add exciton-exciton scattering processes to the dipolariton Hamiltonian and study its influence on dynamics of the system. In particular, we accommodate the model for accounting of the spatial dynamics of a dipolaritonic cloud and define the restrictions on the pumping rates.

As we mentioned before, we consider high pumping rates, which is why the mean field approximation is fully applicable. In the lowest order of mean-field theory, one can treat operators for the modes as classical fields: $\langle \hat{a} \rangle = \Psi_{C}$, $\langle \hat{b} \rangle = \Psi_{DX}$ and $\langle \hat{c} \rangle = \Psi_{IX}$. Then, it is convenient to use the Hamiltonian density of the system, which can be rewritten accounting for non-linear interactions
\begin{align}
\notag \mathcal H(\mathbf{r}) =& \hbar \omega_C |\Psi_{C}|^2 - \frac{\hbar^2}{2 m_{C}} \Psi_{C}^{\dagger}\nabla^2 \Psi_{C} + \hbar \omega_{DX} |\Psi_{DX}|^2 - \frac{\hbar^2}{2 m_{DX}} \Psi_{DX}^{\dagger}\nabla^2 \Psi_{DX} + \hbar \omega_{IX} |\Psi_{IX}|^2 - \frac{\hbar^2}{2 m_{IX}} \Psi_{IX}^{\dagger}\nabla^2 \Psi_{IX} \\ \notag
&+ \frac{\hbar\Omega}{2} (\Psi_{C}^\dagger \Psi_{DX} + \Psi_{DX}^\dagger \Psi_{C}) - \frac{\hbar J}{2}(\Psi_{DX}^\dagger \Psi_{IX} + \Psi_{IX}^\dagger \Psi_{DX}) + \frac{V_{DD}}{2} \Psi_{DX}^\dagger \Psi_{DX}^\dagger \Psi_{DX} \Psi_{DX} \\ &+ \frac{V_{II}}{2} \Psi_{IX}^\dagger \Psi_{IX}^\dagger \Psi_{IX} \Psi_{IX} + V_{DI} \Psi_{DX}^\dagger \Psi_{IX}^\dagger \Psi_{DX} \Psi_{IX} + P(t) \Psi_{C}^\dagger + P(t)^* \Psi_{C},
\label{Hnl}
\end{align}
where $V_{DD}$, $V_{II}$ and $V_{DI}$ are interaction constants describing direct-direct, indirect-indirect and direct-indirect exciton scattering. Here we also introduced kinetic energy terms of cavity photon, direct and indirect excitons with effective masses being $m_{C}$, $m_{DX}$ and $m_{IX}$, respectively. Using Heisenberg equations of motion the classical fields, we can straightforwardly derive dynamical equations for a dipolariton system with the spatial dynamics:
\begin{align}
&i \hbar \frac{\partial \Psi_{C}}{\partial t} = - \frac{\hbar^2 \nabla^2}{2 m_{C}} \Psi_{C} + \frac{\hbar \Omega}{2} \Psi_{DX} + P(t) - i\frac{\Gamma_{C}}{2} \Psi_{C},
\label{Psi_C}\\
&i \hbar \frac{\partial \Psi_{DX}}{\partial t} =  - \frac{\hbar^2 \nabla^2}{2 m_{DX}} \Psi_{DX} + \hbar \delta_\Omega \Psi_{DX} + \frac{\hbar \Omega}{2} \Psi_{C} - \frac{J}{2} \Psi_{IX} + (V_{DD} |\Psi_{DX}|^2 + V_{DI} |\Psi_{IX}|^2) \Psi_{DX} - i\frac{\Gamma_{DX}}{2} \Psi_{DX} ,
\label{Psi_DX}\\
&i \hbar \frac{\partial \Psi_{IX}}{\partial t} =  - \frac{\hbar^2 \nabla^2}{2 m_{IX}} \Psi_{IX} + (\hbar \delta_\Omega - \hbar \delta_J)\Psi_{IX} - \frac{\hbar J}{2} \Psi_{DX} + (V_{II}|\Psi_{IX}|^2 + V_{DI}|\Psi_{DX}|^2)\Psi_{IX} - \frac{\Gamma_{IX}}{2} \Psi_{IX},
\label{Psi_IX}
\end{align}
where we used detuning between modes similarly to main text. Eqns. (\ref{Psi_C})-(\ref{Psi_IX}) represent the system of nonlinear Schr\"{o}dinger equations (or Gross-Pitaevskii equations) for a dipolariton system in $| \Psi \rangle = (\Psi_{C}; \Psi_{DX}; \Psi_{IX})$ basis with introduced decay rates. 
Alternatively, one can switch to the dipolariton basis $| \widetilde{\Psi} \rangle = \mathcal{U} | \Psi \rangle = (\Psi_{UP}; \Psi_{MP}; \Psi_{LP})$ with the Hamiltonian density given by $\widetilde{\mathcal{H}} = \mathcal{U}^{-1} \mathcal{H} \mathcal{U}$. Here $\mathcal{U}$ is a unitary matrix corresponding to Hopfield transformations \cite{KavokinBook}. The system then is described by linear superposition of states, namely: upper, middle and lower dipolaritons.
\begin{figure}[h]
\includegraphics[width=0.8\linewidth]{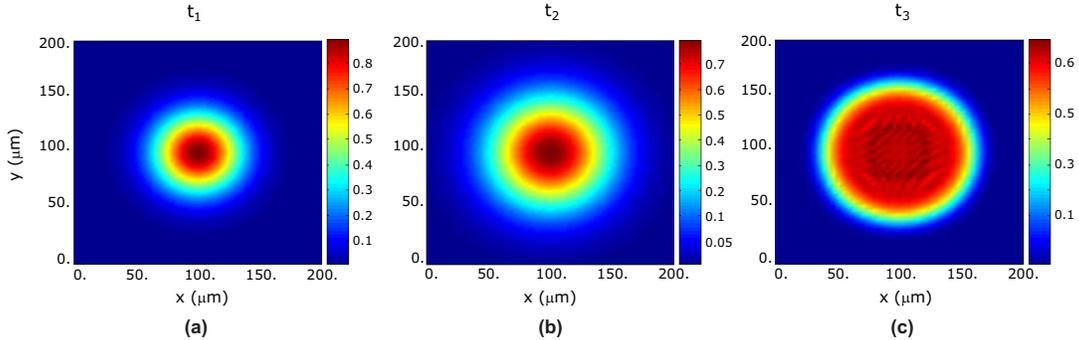}
\caption{(Color online) Spatial dynamics of dipolariton cloud. Simulated density profile is shown for three different time steps: $t_1=0$ ps (a), $t_2=25$ ps (b) and $t_3=60$ ps (c). Horizontal and vertical axes correspond to $x$ and $y$ in-plane Cartesian coordinates, and indirect exciton density is measured in 10$^{10}$ cm$^{-2}$ units.}
\label{FigS2}
\end{figure}

We model the spatial dynamics of the dipolariton system plotting the density of IX component, $|\Psi_{IX}|^{2}$ as a function of time (Fig. \ref{FigS2}). Calculations where performed for 60 $\mu$m diameter of pumping spot, $m_{C}=5\times 10^{-5} m_{e}$ and $m_{DX}=m_{IX}=0.51 m_{e}$ ($m_e$ is free electron mass), and decay times stated before. Interaction constants are estimated as $V_{DD}=4.8~\mu eV~\mu m^2$, $V_{DI}=14~\mu eV~\mu m^2$ and dipole-dipole interaction $V_{II}=39.5~\mu eV~\mu m^2$ \cite{KyriienkoIndirect}, which correspond to $12$ nm separation between QWs. An important timescale of the problem is setted by duration of IX-DX beatings which spans up to 30 ps. We observe that for pumping rate corresponding to indirect exciton density $n_{IX}=10^{10}~cm^{-2}$ the effects of dispersion become pronounced only in 50 ps after pumping pulse. Thus, the effects of interaction do not restrict the prescribed THz emission in the pulsed regime.

However, our initial proposal is not restricted to Gaussian profile pump and can be easily performed with homogeneous pumping in real space. Additionally, the system can be placed to a potential well or micropillar cavity, where the issue of spatial dispersion is insignificant.

\subsection{D: Dynamics analysis}
In the work we have studied the dynamics of indirect exciton density oscillations which induce emission of coherent THz radiation. These oscillations emerge due to the strong coupling between three modes, namely: indirect excitons (IX), direct excitons (DX) and cavity photons (C). In general, the dynamics of dipolaritons described by system of coupled equations for occupation numbers and correlators is complex due to the presence of mixing terms, detuning, decay and tunable pumping conditions. It can be described by characteristic energies $\hbar J$ and $\hbar \Omega $ which denote IX-DX and C-DX coupling, respectively, detuning parameters $\delta _{J}=\omega _{IX}-\omega _{DX}$ and $\delta_{\Omega }=\omega_{C}-\omega _{DX}$, and pumping frequency detuning from the cavity mode $\Delta =\omega_{P}-\omega_{C}$. Consequently, in order to comprehend the dynamics of the system one needs to study the dependence of density oscillations on five independent parameters, which is a non-trivial task.
\begin{figure}
\includegraphics[width=0.85\linewidth]{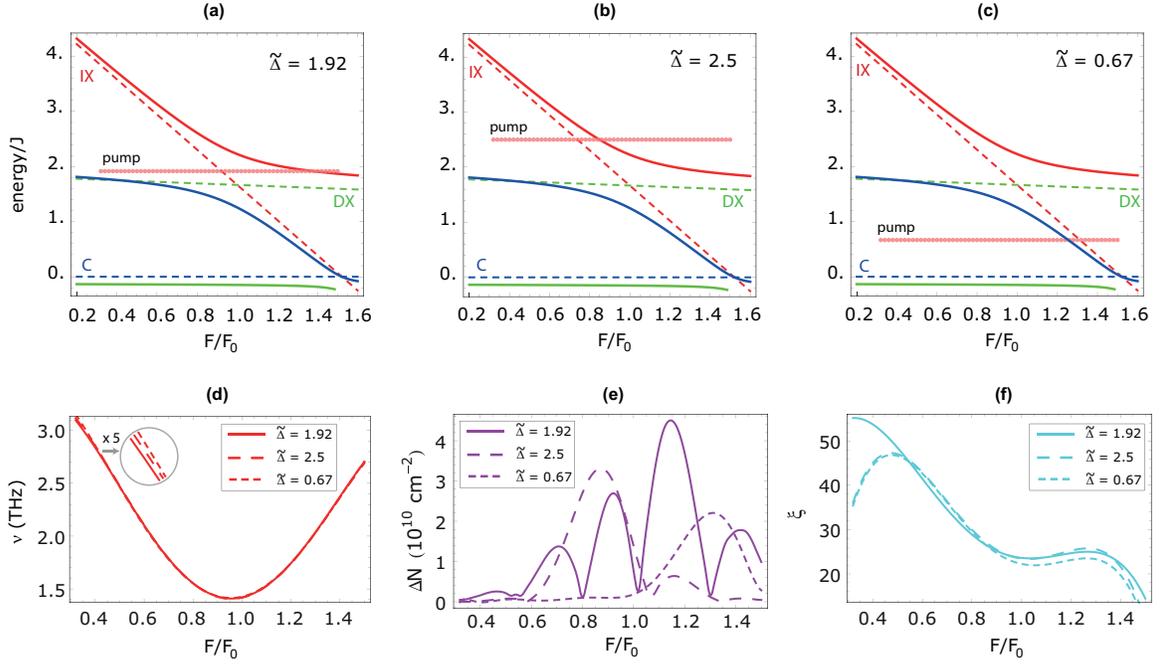}
\caption{(Color online) (a), (b) and (c): Energy spectrum of dipolariton modes (solid lines) as a function of normalized electric field $F/F_{0}$ plotted for $\delta_{\Omega }=-1.67$. We assume equal IX-DX and C-DX coupling constants, measure energy in units of $J$ and choose cavity mode as a reference point, $E_{C}=0$. Pink dotted lines show energy position
of the pumping of the cavity mode. (d) Frequency of long-term indirect exciton density oscillations $\nu$ for different electric field, showing the possible tunable window of the signal frequency. Tunnelling frequency used for calculations is $J/2\protect\pi =1.45$ THz. (e) Amplitude of oscillations showing the correspondence between pumping conditions and
mode detuning. (e) Quality factor of the oscillations defined as ratio between amplitude and decrement of oscillation.}
\label{FigS3}
\end{figure}

The coupling constants of light-matter interaction $\hbar \Omega$ and indirect exciton to direct exciton coupling $\hbar J$ are intrinsic characteristics of the structure. The structure can be designed in such a way that these two characteristics coincide: $\hbar \Omega =\hbar J$, in which case DX, IX and C modes are efficiently intermixed. This regime has already been studied in the samples where dipolariton modes have been observed for the first time \cite{Cristofolini,ChristmannAPL} with $\hbar \Omega =\hbar J=6$ meV. Therefore, the behavior of dipolaritons can be controlled tuning the energy spacing between modes by altering the applied electric field ($\delta_{J}$) and tuning the incidence angle of the excitation light ($\delta _{\Omega}$).

We shall specifically consider two particular cases. In the first case the cavity mode is far detuned from the IX-DX resonance, which partially resembles the weak light-matter coupling (regime I). The second case refers to the resonance of all three modes achieved for a certain incidence angle where $\delta_{\Omega}$ is small (regime II). For each regime we
study the dependence of properties of indirect exciton density oscillations on the applied voltage. To keep the results general, the analysis is performed in dimensionless units $F/F_{0}$, where electric field is normalized to value where IX and DX resonances anti-cross, $F_{0}$. Moreover, we measure the frequency in units of tunnelling coupling $J$. Finally, the last parameter which governs the dynamics of the system is the pumping frequency $\omega_{P}=\Delta +\omega_{C}$.
\begin{figure}
\includegraphics[width=0.85\linewidth]{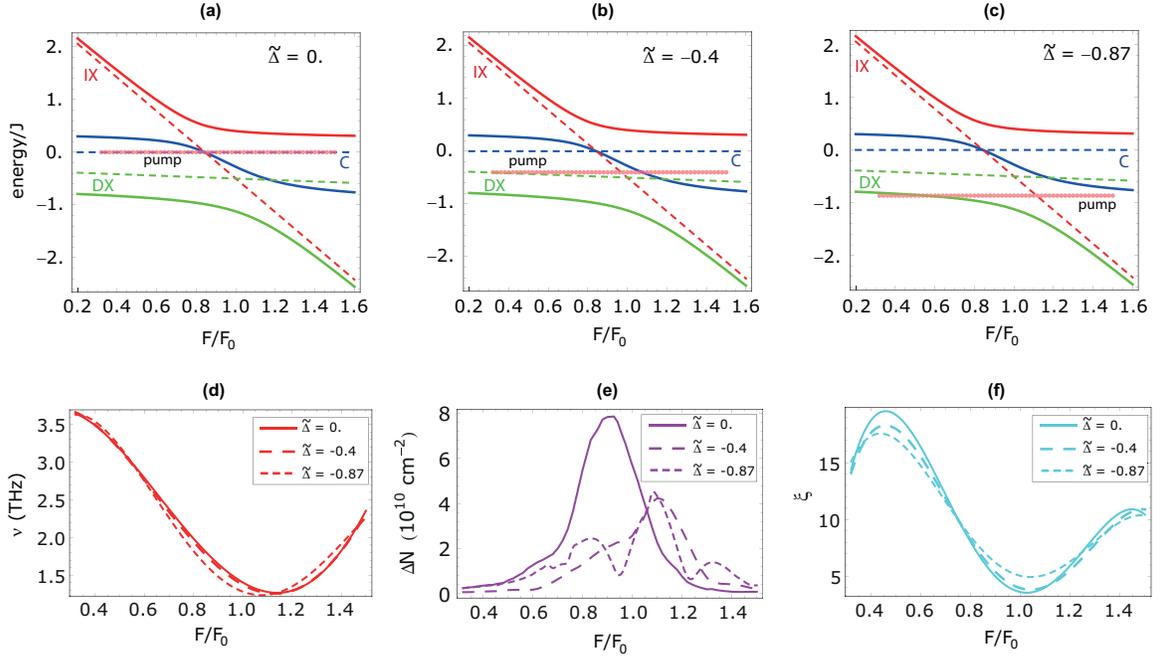}
\caption{(Color online) (a), (b) and (c): Energy spectrum of dipolariton modes (solid lines) as a function of normalized electric field $F/F_{0}$ plotted for $\delta_{\Omega}=0.5$. Pink dotted lines show energy position of the pumping of the cavity mode. (d) Frequency of long-term indirect exciton density oscillations $\nu$ for different electric field, showing the range where signal frequency can be tuned. (e) Amplitude of oscillations showing the correspondence between pumping conditions and mode detuning. (e) Quality factor of the oscillations defined as ratio between amplitude and decrement of oscillation.}
\label{FigS4}
\end{figure}

In the main text of this paper we present the plot of indirect indirect exciton density oscillations for two particular situations [Fig. 2 in the main text]. In general case the dynamics exhibits turning-on transitional region lasting for several picoseconds after the arrival of the pulse followed by subsequent long-term oscillations of density decaying in time. Here we vary the pumping conditions and study the \textquotedblleft quality\textquotedblright\ of oscillations. To do this we use the following set of parameters. Complementary to the frequency of long-range oscillations, we find the averaged amplitude of the oscillations in transient regime $\Delta N$ to define the efficiency of pumping conditions. This quantity strongly affects the power output of generated initial THz signal and it is important for achievement of powerful pulsed THz emission. Additionally, we define dimensionless oscillation quality factor $\xi$ defined as a ratio of long-term oscillation amplitude $\Delta N$ to its decrement in time $\delta N$. Therefore, high $\xi $ stands for slowly decaying oscillations with comparably large amplitude.

We consider the first regime when cavity mode is far-detuned from the IX-DX anti-crossing region, $\delta_{\Omega }=-1.67$. In energy diagrams in Fig. \ref{FigS3}(a,b,c) we chose three pumping energies corresponding to the resonant pumping ($\Delta =1.92$), blue detuned pumping ($\Delta =2.5$) and red-detuned pumping ($\Delta =-0.67$). Solid lines there denote upper, middle and lower dipolariton branches (red, blue and green lines, respectively). The energy is counted from the bottom of the cavity mode ($E_{C}=0$). Fig. \ref{FigS3}(d) shows dependence of the frequency of $n_{IX}$ oscillations as function of the electric field for different pumping energy shown in energy diagrams in Fig. \ref{FigS3}(a,b,c) (pink line). One can see that it has the minimum value for almost resonant electric field $F/F_{0}\approx 0.9$ and approximately corresponds to tunnelling coupling $\nu \approx J/2\pi = 1.45$ THz. Also, it increases with the increase of detuning between IX and DX modes, $\delta _{J}$. Moreover, we note that the frequency of stabilized oscillations does not depend on the pumping conditions. In Fig. \ref{FigS3}(e) we show amplitude of oscillations. It differs drastically for three pumping frequencies. One can see that $\Delta N$ increases if pump coincides with energy bare transition or dipolariton mode, and this correlation can be tracked in Fig. \ref{FigS3}(e) for all detunings. However, calculations show that oscillations with the largest amplitude correspond to the strongest decrement of oscillations, which result into minimal $\xi $ in the resonant region [Fig. \ref{FigS3}(f)]. Out of the resonance both amplitude oscillation and decay are small, and the realistic region should be chosen between electric field corresponding to maxima of the quality factor $\xi$.

Next, we study the second regime where cavity mode is detuned from direct exciton by $\delta_{\Omega}=0.5$. In the IX-DX resonance point dipolariton represents strong mixture of bare modes with two anti-crossings shown in Fig. \ref{FigS3}(a-c). The dynamics of the system was calculated for three pumping energies which change from resonance [Fig. \ref{FigS3}(a), $\Delta=0$] to out-of-resonance conditions [Fig. \ref{FigS3}(c), $\Delta=-0.87$]. Fig. \ref{FigS4}(d) shows no influence of pumping conditions on the oscillation frequency, while the position of the minimum shifts to $F/F_{0}\approx 1.1$ point. An amplitude of oscillations in Fig. \ref{FigS4}(e) shows the same correlation with respect to pumping conditions observed in the first regime, while its absolute value is larger. However, the quality factor $\xi$ of oscillation in second regime is several times lower due to faster decay of oscillations [Fig. \ref{FigS4}(f)].

Thus, we can conclude that the first regime is appropriate for the cases where stable long-term emission of THz radiation is required, while the second regime is needed for the pulsed generation of powerful signals.

\subsection{E: Sketch of system with supplemental THz cavity}
\begin{figure}[h!]
\includegraphics[width=0.4\linewidth]{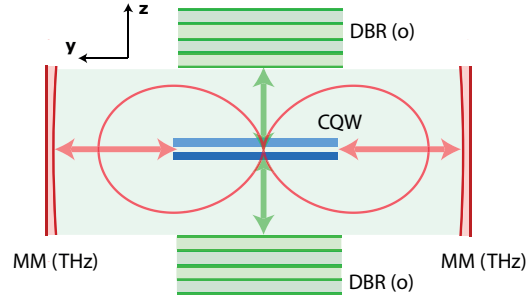}
\caption{(Color online) Sketch of the geometry where power of THz emission is increased due to Purcell effect (top view). Here DBR (o) represent optical cavity mirrors (green), while MM (THz) denote metallic mirror cavity tuned to reflect signal of THz frequency (red). The red lines show peculiar polar pattern of dipolariton THz emitter.}
\label{FigS5}
\end{figure}

\end{widetext}


\begin{thebibliography}{99}

\bibitem{Siegel} P. H. Siegel, IEEE Trans. Microw. Theory Techn. \textbf{50}, 910 (2002).

\bibitem{Gallerano} G. P. Gallerano and S. Biedron, in \textit{Proceedings of the 2004 FEL Conference, Trieste, Italy, 2004}, p. 216-221.

\bibitem{Eisele} H. Eisele, IEEE Trans. Microw. Theory Techn. \textbf{48}, 626 (2000).

\bibitem{Faist} J. Faist, F. Capasso, D. L. Sivco, C. Sirtori, A. L. Hutchinson, and A. Y. Cho, Science \textbf{264}, 553 (1994).

\bibitem{Kohler} R. K\"ohler {\it et al.}, Nature \textbf{417}, 156 (2002).

\bibitem{Geiser} M. Geiser, F. Castellano, G. Scalari, M. Beck, L. Nevou, and J. Faist, Phys. Rev. Lett. \textbf{108}, 106402 (2012).

\bibitem{Auston} D. H. Auston, K. P. Cheung, J. A. Valdmanis, and D. A. Kleinman, Phys. Rev. Lett. \textbf{53}, 1555 (1984).

\bibitem{Fattinger} C. Fattinger and D. Grischkowsky, Appl. Phys. Lett. \textbf{54}, 490 (1989).

\bibitem{Heinz} J. Shan and T. F. Heinz, Topics Appl. Phys. \textbf{92}, 59 (2004).

\bibitem{Johnston} M. B. Johnston, D. M. Whittaker, A. Corchia, A. G. Davies, and E. H. Linfield, Phys. Rev. B \textbf{65}, 165301 (2002).

\bibitem{Kibis} O. V. Kibis, M. R. da Costa, and M. E. Portnoi, Nano Lett. \textbf{7}, 3414 (2007).

\bibitem{Portnoi} K. G. Batrakov, O. V. Kibis, P. P. Kuzhir, M. R. da Costa, and M. E. Portnoi, J. Nanophoton. \textbf{4}, 041665 (2010).

\bibitem{Alexeev} A. M. Alexeev and M. E. Portnoi, Phys. Rev. B \textbf{85}, 245419 (2012).

\bibitem{KavokinBook} A. V. Kavokin, J. J. Baumberg, G. Malpuech, and F. P. Laussy, \textit{Microcavities} (Oxford University Press, Oxford, 2007).

\bibitem{PolaritonDevices} T. C. H. Liew, I. A. Shelykh, and G. Malpuech, Physica E \textbf{43}, 1543 (2011).

\bibitem{KVKavokin} K. V. Kavokin, M. A. Kaliteevski, R. A. Abram, A. V. Kavokin, S. Sharkova, and I. A. Shelykh, Appl. Phys. Lett. \textbf{97}, 201111 (2010).

\bibitem{Savenko} I. G. Savenko, I. A. Shelykh, and M. A. Kaliteevski, Phys. Rev. Lett. \textbf{107}, 027401 (2011).

\bibitem{delValle} E. del Valle and A. V. Kavokin, Phys. Rev. B \textbf{83}, 193303 (2011).

\bibitem{Kavokin} A. V. Kavokin, I. A. Shelykh, T. Taylor, and M. M. Glazov, Phys. Rev. Lett. \textbf{108}, 197401 (2012).

\bibitem{Liew} T. C. H. Liew, M. M. Glazov, K. V. Kavokin, I. A. Shelykh, M. A. Kaliteevski, and A. V. Kavokin, Phys. Rev. Lett. \textbf{110}, 047402 (2013).

\bibitem{Cristofolini} P. Cristofolini, G. Christmann, S. I. Tsintzos, G. Deligeorgis, G. Konstantinidis, Z. Hatzopoulos, P. G. Savvidis, and J. J. Baumberg, Science \textbf{336}, 704 (2012).

\bibitem{Lozovik} Yu. E. Lozovik and V. I. Yudson, Sov. Phys. JETP \textbf{44}, 389 (1976).

\bibitem{ButovPRL} L. V. Butov {\it et al.}, Phys. Rev. Lett. \textbf{86}, 5608 (2001).

\bibitem{SnokeScience} D. Snoke, Science \textbf{298}, 1368 (2002).

\bibitem{ButovNature} A. A. High, J. R. Leonard, A. T. Hammack, M. M. Fogler, L. V. Butov, A. V. Kavokin, K. L. Campman, and A. C. Gossard, Nature \textbf{483}, 584 (2012).

\bibitem{KyriienkoIndirect} O. Kyriienko, E. B. Magnusson, and I. A. Shelykh, Phys. Rev. B \textbf{86}, 115324 (2012).

\bibitem{ChristmannPRB} G. Christmann, C. Coulson, J. J. Baumberg, N. T. Pelekanos, Z. Hatzopoulos, S. I. Tsintzos, and P. G. Savvidis, Phys. Rev. B \textbf{82}, 113308 (2010).

\bibitem{ChristmannAPL} G. Christmann, A. Askitopoulos, G. Deligeorgis, Z. Hatzopoulos, S. I. Tsintzos, P. G. Savvidis, and J. J. Baumberg, Appl. Phys. Lett. \textbf{98}, 081111 (2011).

\bibitem{Kardiff} K. Sivalertporn, L. Mouchliadis, A. L. Ivanov, R. Philp, and E. A. Muljarov, Phys. Rev. B \textbf{85}, 045207 (2012).

\bibitem{Bayer} M. Bayer, V. B. Timofeev, F. Faller, T. Gutbrod, and A. Forchel, Phys. Rev. B \textbf{54}, 8799 (1996).

\bibitem{Kyriienko2011} O. Kyriienko and I. A. Shelykh, Phys. Rev. B \textbf{84}, 125313 (2011).

\bibitem{SM} See supplemental material at ``weblink''.

\bibitem{ButovJPCM} L. V. Butov, J. Phys.: Condens. Matter \textbf{16}, R1577 (2004).

\bibitem{Landau} L. D. Landau and E. M. Lifshitz, \textit{The Classical Theory of Fields} (Butterworth-Heinemann, 1980).

\bibitem{Dicke} R. H. Dicke, Phys. Rev. \textbf{93}, 99 (1954).

\bibitem{Bohnet} J. G. Bohnet, Z. Chen, J. M. Weiner, D. Meiser, M. J. Holland, and J. K. Thompson, Nature \textbf{484}, 78 (2012).

\bibitem{Yukalov} V. I. Yukalov and E. P. Yukalova, Phys. Rev. B \textbf{81}, 075308 (2010).

\bibitem{Walther} C. Walther, G. Scalari, M. I. Amanti, M. Beck, and J. Faist, Science \textbf{327}, 1495 (2010).

\end{thebibliography}

\begin{thebibliography}{99}

\bibitem{Carmichael} H. Carmichael, \textit{An open systems approach to quantum optics} (Springer, Germany, 1993).

\bibitem{Schlosshauer} M. A. Schlosshauer, \textit{Decoherence and the Quantum-to-Classical Transition} (Springer, Berlin, 2007).

\bibitem{Laussy} A. Gonzalez-Tudela, E. Del Valle, E. Cancellieri, C. Tejedor, D. Sanvitto, F. P. Laussy, Optics Express \textbf{18}, 7002 (2010).

\bibitem{HP} T. Holstein and H. Primakoff, Phys. Rev. \textbf{58}, 1098 (1940). 

\end{thebibliography}
\end{document}